\documentclass[11pt]{article}

\usepackage[margin=1in]{geometry}

\usepackage[T1]{fontenc}
\usepackage{lmodern}

\usepackage{setspace}
\usepackage{amsmath}
\usepackage{amssymb}
\usepackage{amsthm}
\usepackage{bbm}
\usepackage{scalerel}

\usepackage{graphicx}
\usepackage{tikz}
\usepackage{adjustbox}

\usepackage{array}
\usepackage{multirow}

\newcolumntype{C}[1]{>{\centering\arraybackslash}p{#1}}

\usepackage{algorithm}
\usepackage{algorithmic}

\usepackage{natbib}

\usepackage{url}
\usepackage{enumerate}
\usepackage{placeins}
\usepackage{verbatim}

\usepackage[
    colorlinks=true,
    citecolor=blue,
    linkcolor=blue,
    urlcolor=blue
]{hyperref}

\theoremstyle{plain}

\theoremstyle{definition}

\theoremstyle{remark}

\numberwithin{equation}{section}
\usepackage{setspace}
\title{
A Bayesian framework for multilevel data under model mis-specification
}

\author{
Widemberg S. Nobre$^{1}$,
David A. Stephens$^{2}$,
Alexandra M. Schmidt$^{3}$,
and \\ Erica E. M. Moodie$^{3}$
\\[6pt]
\small
$^{1}$Departamento de M\'etodos Estat\'isticos,
Universidade Federal do Rio de Janeiro, Brazil
\\
\small
$^{2}$Department of Mathematics and Statistics,
McGill University, Canada
\\
\small
$^{3}$Department of Epidemiology and Biostatistics,
McGill University, Canada
\\[6pt]
\small
Corresponding author: Widemberg S. Nobre
\\
\small
\texttt{widemberg@im.ufrj.br}
}

\date{}

\begin{document}

\maketitle

\begin{abstract}
We propose a Bayesian framework for uncertainty quantification from the perspective that the working model is mis-specified in settings of a multilevel data-generating process. We focus on settings in which the mis-specification fails to match the functional form of the mean structure, and discuss Bayesian estimation of target parameters under dependence induced by a mismatch between working and data-generating models. The proposal represents a Bayesian semi-parametric procedure aimed at estimating population-level parameters while accounting for cluster- and unit-level variation in the estimating function. The proposal extends the regular Bayesian bootstrap to account for cluster- and unit-level variation using multilevel weights from an enriched Dirichlet model. Simulation studies indicate that the proposed approach has good frequentist properties when the data-generating process and the proposed model induce a partially exchangeable sequence associated with the unknown quantity of interest. Applications to radon \citep{gelman2007data}, Programme for International Student Assessment 2022 \citep{OECD2023PISA}, and tuberculosis \citep{nobre2023impact} datasets are presented for illustrative purposes. The results demonstrate that the proposed method is competitive with variations of multilevel models, with major differences observed in the range of credible intervals, which are justified by the nonparametric assumptions underlying the proposed method.
\end{abstract}


\bigskip
	
	\section{Introduction}\label{sec:intro}
	
	Bayesian inference relies on translating the law of probability into uncertainty quantification, and usually concerns updating prior to posterior beliefs through the application of Bayes' theorem. 	Given the challenges and advances of statistical analysis, some of the current \textit{Bayesianism} culture introduced by \cite{deFinetti1974,deFinetti1975} and \cite{savage1972foundations} has been amended in different ways. Such amendments are often motivated by complexity and the target of interest, although this may not always be true -- see \cite{gelman2008objections} and its discussion for more details. 
	Here, we term \textit{Bayesianism} as illustrative of a theory that adheres to three pillars: prior-to-posterior beliefs, uncertainty quantification via probabilistic arguments, and parameter interpretation, which are direct interpretations following from de Finetti's representation theorem. However, we argue that, in some sense, a coherent Bayesian specification may raise controversy, particularly when deliberately assuming mis-specified models.	Mis-specification in Bayesian inference has been explored extensively albeit in limited settings.  In this paper, we address the issue of mis-specification in multilevel models.
	
    {From a parametric perspective, Bayesian mis-specified models are characterized by either the set of probability measures implied by the likelihood model not containing the true data-generating process, or by the true parameter value not lying in the prior support.}  In such cases, model mis-specification may render Bayes' theorem inappropriate for learning parameters, motivating decision-theoretic approaches to implement Bayesian inference. In this regard, one may handle model mis-specification by targeting a `guess' at the unknown parameter which is the best in the sense of minimizing the Kullback–Leibler divergence between the proposed model and the true/target distribution. From a Bayesian perspective, we do agree that such a \textit{best guess parameter} has a clear interpretation as demonstrated in \cite{walker2013bayesian} and \cite{de2013bayesian,10.1214/13-BA024}, although it may fail to address particular settings of interest. {For instance, in causal inference, the goal is typically to describe the identifiability assumptions under a given setting of interest and the proposed method. Thus, an unknown best guess may be incoherent without further guarantees.} Specifically, the identification of causal parameters requires additional structural assumptions, which make such a \textit{guess} causally interpretable under certain conditions.

	When discussing conscious model mis-specification, \cite{10.1214/22-BA1322} relies on nonparametric arguments to justify the use of the Bayesian bootstrap \citep{rubin1981Bayesian,newton1994approximate} for correct inference. The argument is made under the assumption of exchangeable sequences for observables, whereas the conscious mis-specification follows from propensity score regression \citep{rosenbaum1983central}. In their proposed method, the propensity score function is considered as a randomization probability by making the strong ignorability assumption \citep{rubin1974estimating}. Here, we extend these ideas to the setting of multilevel data \citep{gelfand2007multilevel}, where observations are nested within clusters, and both the number of clusters and the cluster to which each unit belongs are known. 
    
    {We aim to simplify model design and retain the computational tractability of simpler models relative to more general hierarchical formulations, and to accommodate conscious model mis-specification while targeting correct uncertainty quantification. In this regard, our proposal extends the semi-parametric formulation presented in \cite{10.1214/22-BA1322} by considering weights structured according to an enriched Dirichlet process \citep{wade2011enriched,quintana2022dependent}.} Indeed, the contribution can be seen as an extension of the Bayesian bootstrap to address multiple levels of uncertainty (individual and cluster) in the objective function, in light of mis-specified models.
	
	Several extensions of the traditional Bayesian bootstrap have been proposed. Some recent references include \cite{10.1214/19-BA1155} and \cite{10.1214/17-BA1071}, which are motivated by massive data and an intractable likelihood, respectively. \cite{oganisianBIO} proposes a hierarchical Bayesian bootstrap that shares covariate information across strata in order to estimate heterogeneous treatment effects. Another interesting variant of the Bayesian bootstrap is proposed in \cite{cui2024bayesian}, which extends the Bayesian bootstrap to mixture distributions. The method relies on the concept of Martingales \citep{fong2023martingale} by exploring the relationship between the Pólya-urn model and the traditional Bayesian bootstrap, while adapting it as a stochastic gradient-type algorithm. 
	
	The fundamental objective of this paper is to develop a Bayesian bootstrap scheme that can handle hierarchical data, even when the number of hierarchies in the proposed model is mis-specified. At the same time, the method should provide reliable uncertainty quantification in the estimation of target parameters. In this regard, Bootstrap methods for multilevel data have been investigated in the frequentist setting. \cite{field2007bootstrapping} compare the first two moments of sums of squares statistics and shows that, in a simple random (cluster) intercept model, the variances and covariances of the between- and within-cluster sums of squares statistics differ depending on whether the model is specified with random effects or marginalized (integrating out the random effect). The authors also state conditions under which consistency is achieved for each of the bootstrap approaches considered; an extension to robust estimators is provided in \cite{field2010bootstrapping}. From a Bayesian perspective, \cite{newton1994approximate} discuss weighted bootstrap methods for dependent data by proposing a special case of the blockwise bootstrap, which selects blocks of consecutive observations in a time-series setting \citep{kunsch1989jackknife,buhlmann1994blockwise}. It is worth noting that similarities between Dirichlet and Multinomial distributions may appeal to adapting frequentist arguments into a Bayesian perspective, which may translate into first-order asymptotic equivalence \citep{rubin1981Bayesian,newton1994approximate}. 
	{In this paper, we focus on estimating population-level parameters. The contributions in this direction are two-fold. Firstly, we propose Bayesian justifications for populational-level estimands in terms of exchangeable and partial exchangeable sequences. Second, we provide an extension of the proposal in \cite{10.1214/22-BA1322} that accommodates the estimation of population-level parameters under multilevel structure.} 
	
	The remainder of this paper is organized as follows. Section \ref{sec:bf} presents the setup of interest and discusses the adequacy of the proposed method of \cite{10.1214/22-BA1322} when handling multi-level data. Additionally, it also includes a nonparametric justification for the proposed method under partial exchangeability. The proposed method is illustrated in Section \ref{sec:illust} through simulation studies and real data applications. We conclude with a discussion in Section \ref{sec:disc}.

	\section{General Bayesian Framework and Proposed Method}\label{sec:bf}

	Let $({O})_{ji}=(Y,Z,X)_{ji}$, for $j=1,\cdots,M$ and $i=1,2,\cdots,n_j$, be a set of observable variables, with $y_{ji}$ and $z_{ji}$ denoting, respectively, outcome and exposure of interest for unit $i$ at cluster $j$, and $X_{ji}$ a set of observed covariates. The general scope of this paper presumes that a set of observations is constructed based on the following particular decomposition of the joint probability model, 
	\begin{equation}
		p_{X,Z,Y}(x,z,y) = p_X(x)p_{Z|X}(z|x)p_{Y|Z,X}(y|z,x),
		\label{eq:joint_obs_model}
	\end{equation}
	which is justified regarding a temporality assumption, i.e., the variable $X$ occurs prior to $Z$ and $Y$, whereas $Z$ is observed prior to $Y$. 
	
	Consider that a set of observed data is available. Define $\mathbf{O}_{j} = ({O}_{j1},\cdots,{O}_{jn_j})^\top$, and let $\mathbf{O}_{1:M}$ denote the stacked vector formed by $\mathbf{O}_{1}$, $\mathbf{O}_{2}$, $\cdots$, $\mathbf{O}_{M}$. The joint probability density for the observed data is given by
	$p\left(\mathbf{o}_{1:M}\right) = p\left(\mathbf{o}_{1}, \cdots, \mathbf{o}_{M}\right)$.  In our setup, it is appealing to consider a joint distribution that is symmetric with respect to permutations of the indices $\{1,\cdots,M\}$ and also within each index $j\in \{1,\cdots,M\}$; we refer to this structure as a partially exchangeable sequence with multilevel structure. A more restrictive assumption is considered in \cite{10.1214/22-BA1322}, where each term on the right side of the decomposition in \eqref{eq:joint_obs_model} is assumed to be invariant to all permutations of indices. The estimand of interest for this paper is defined as
	\begin{equation}
		\begin{split}
			\tau &= \int \psi(x,z,y) d F_O(x,z,y;\theta_0),
		\end{split}
		\label{eq:target_par}
	\end{equation}
	where $F_O (\cdot;\theta_0)$ denotes the true data-generating distribution. However, we also assume that inference is carried out under an alternative model, say $F (\cdot;\theta)$, and the goal is to produce correct inference for the target parameter $\tau$.

\subsection{Posterior sampling of target parameters: exchangeability \textit{vs} partial exchangeability}\label{sec:cexc}

Following \cite{10.1214/22-BA1322}, let the sequence of observations be independent given some true distribution $F_O(\cdot; \theta_0)$. This understanding is consistent with the de Finetti's representation theorem under exchangeability. Within this framework, the authors adopt a decision-theoretic approach for inference, which leads to an inference procedure that ultimately applies the Bayesian bootstrap.

The inference procedure follows through the maximization of a utility function $u_\theta(\cdot,\cdot)$. Defining $u_\theta(s,t^\prime)$ as the Kullback-Leibler divergence from $F_O (s;\theta)$ to some proposed model $F (s;t^\prime)$, we may obtain posterior samples from a target parameter $\vartheta$ by performing the maximization,
\[
\vartheta = arg\max_{t^\prime \in \Theta^\prime} \int \left\{\int \log f(s;t^\prime) f_O(s;\theta)ds \right\}\pi_N(\theta)d\theta,
\] 
where $\pi_N(\cdot)$ denotes the posterior distribution constructed via Bayes' theorem based on the correct model, and $\Theta^\prime$ the parameter space, with $N = \sum_{j=1}^{M}  n_j$. If a single posterior sample $\theta^{(l)}$ is available from $\pi_N(\cdot)$, then a posterior sample $\vartheta^{(l)}$ is available by the transformation stated in the inner integral with respect to $s$ and conditional on $\theta^{(l)}$. 

Consider, however, that $f_O(s;\theta^{(l)})$ is not available; it follows that the integral over $s$ is analytically intractable. As a remedy for this setting, the resulting integral can be approximated using Monte Carlo integration since the units $S_{11},\cdots,S_{1n_1},S_{21},\cdots,S_{Mn_M}$ comprise independent samples from $F_O$. 
Alternatively, one may posit a symmetric Dirichlet process prior on $F_O$, in a procedure that ultimately translates into the Bayesian bootstrap \citep{10.1214/22-BA1322}, with concentration parameter $\zeta \rightarrow 0$ in the limiting case. As a consequence, the estimation procedure is such that
 \[
 \vartheta = \arg \max_{t^\prime \in \Theta^\prime} \sum_{j=1}^M \sum_{i=1}^{n_j}\omega_{ji} \log f(s_{ji};t^\prime),
 \] 
 where $(\omega_{11},\cdots,\omega_{1n_1},\omega_{21},\cdots,\omega_{Mn_M})\sim Dirichlet (1,\cdots,1)$, which reflects the regular Bayesian bootstrap with a weighted likelihood \citep{newton1994approximate}. 
	
A critical consideration is whether the conditional independence assumption is satisfied. In this regard, suppose in Equation \eqref{eq:target_par} that the goal is to estimate $\tau = E(\psi)$, with $\psi$ introduced from the data-generating model and proposed model such that a sequence $[\psi_{ji}]_{j = 1,\cdots,M;i=1,\cdots,n_j}$ is deterministic in terms of a parametric model made over the set of observable quantities. We seek a nonparametric characterization of the problem that attempts to accommodate the property,
\begin{equation*}
	\psi |\tilde{P} \overset{ind}{\sim} \tilde{P},\;\text{for $\tilde{P}$ some sampling distribution.}
	\label{eq:psi}
\end{equation*}
Specifically, we assume that $\tilde{P}$ is a probability measure that is a deterministic transformation relating the true data-generating mechanism and the proposed parametric model design. Thus, for correct uncertainty quantification, $\tilde{P}$ is taken to carry all sources of variation in order to be consistent with the conditional independence in the nonparametric formulation; this interpretation is what motivates the setting of random sequences with multilevel structure (see Section \ref{sec:MSOPE} of the supplementary material for details).

	\subsection{A Bayesian bootstrap with enriched Dirichlet weights} \label{sec:bbhs}
	
	Let $F(\cdot|\theta)$ be a probability measure for the proposed model, parameterized by $\theta$. By the law of iterated expectations, we can rewrite the target parameter in \eqref{eq:target_par} as follows:
	\begin{equation}
	\tau =   \int \psi(x,z,y) d F_O(x,z,y;\theta_0)= \int \left[\int \psi(x,z,y)d F(x,z,y;\theta)\right] d F_O(x,z,y;\theta_0),
    \label{eq:lawitexp}
	\end{equation}
	with $F$ and $F_O$ defined in the same probability space. This expression suggests that the inner expectation, taken with respect to $F(\cdot;\theta)$, represents an average under $F_O(\cdot;\theta_0)$. Often, $F(\cdot;\theta)$ is such that this inner expectation can be computed analytically. 
	
	For instance, one may be interested in the average treatment effect (ATE), which attempts the inner expectation in \eqref{eq:lawitexp}, with respect to $F$, be defined as
	\[
	\psi(X,Z,Y;\theta) = E(Y|Z=z+1,X;\theta) - E(Y|Z=z,X;\theta)=\psi(X;\theta),
	\]
	which is the conditional impact of a unit increase in $Z$ on $Y$, for $z \in \mathbb{R}$; alternatively, $\psi$ may also represent an associational parameter between $Z$ and $Y$ derived from the design model $Y|Z,X$. 
	 
	For a given pair of indices $(j,i)$, the parameter of interest may be expressed conditionally on $X$ as
	\begin{equation*}
		\begin{split}
			\tau & =   \int_{\mathcal{X}} \psi(X_{ji}=x;\theta) F_{O}(dx),
		\end{split}
	\end{equation*}
	where the evaluation of $\psi(X_{ji};\theta)$ is embedded into model adjustment, for all $j$ and $i$. Under mild assumptions, we may use the transformation representation of the expected value and rewrite the above integral in terms of
	\[
	\int_{\mathcal{X}} \psi(X_{ji}=x;\theta) F_{O}(dx) = \displaystyle \int_{ \mathcal{A}} \tilde{\psi} \tilde{P}(d\tilde{\psi}),
	\]
	where $\mathcal{A}$ is the set of inverse point mapping and $\tilde{P}$ is some probability measure resulting from the given transformation, which is related to $X$-space. If the observed data and model adjustment match the random sequence ${\psi(X_{ji};\theta)}_{j=1\cdots,M, i=1,\cdots,n_j}$ such that $\psi(X;\theta)\mid \tilde{P}  \overset{ind.}{\sim} \tilde{P}$, then the properties of the proposed method described in Section \ref{sec:cexc} follows as usual, but the same might not be said when the resultant sequence exhibits dependence within or across clusters.
	
	In this regard, our proposal proceeds by assuming that a conditional independent sequence is achieved by approaching $\tilde{P}$ in a product space $({\tilde{\mathcal{X}}\times \tilde{\mathcal{U}}},{{\mathcal{B}}_X\times\mathcal{B}_U})$ where $\tilde{\mathcal{X}}$ and $\tilde{\mathcal{U}}$ are complete and separable metric spaces with ${\mathcal{B}}_X$ and ${\mathcal{B}}_U$ their respective Borel $\sigma$-algebras. Further, we also consider probability measures that are jointly measurable in $(\tilde{P}_U,\tilde{P}_{X|U})$, with $\tilde{P}_U$ and $\tilde{P}_{X|U}$ exchangeable probability measures, and independent between them; this approach is consistent with the partially exchangeable sequence with multilevel structure presented in Section \ref{sec:MSOPE} of the supplementary material. Under such formulation, we may rewrite
	\begin{equation}
		\begin{split}
			\int_{ \mathcal{A}} \tilde{\psi} \tilde{P}(d\tilde{\psi}) & = \iint_{\tilde{\mathcal{X}} \times \tilde{\mathcal{U}}} \left[  \psi( x,u) \tilde{P}_{X|U}(dx|u)\right]\tilde{P}_U(du).
		\end{split}
		\label{eq:ate3_par}
	\end{equation}
	The sequence $\{\psi(\theta;X_{ji},U_{j})\}_{j=1\cdots,M, i=1,\cdots,n_j}$ exists in light of the observed dataset and model adjustment; indeed, such a sequence must be deterministically expressed given data and the proposed model. Assuming that the conditional measures $\tilde{P}_{X|U}$ and the marginal measure $\tilde{P}_U$ are independent Dirichlet processes, and once data are observed, the integral in \eqref{eq:ate3_par} can be approximated through successive applications of Dirichlet mixture weights for all conditional and marginal measures. If the marginal and conditional independence laws along the sequence of $\psi$’s are satisfied, then the solution is exact as the number of clusters and the number of units within each cluster both go to infinity \citep{wade2011enriched}. 
	
	Finally, it is straightforward to show that a decision-theoretic approach can be implemented, similarly to \cite{10.1214/22-BA1322} and inspired by the Bayesian bootstrap, to produce posterior samples for target parameters. Thus, considering the same construction of Section \ref{sec:cexc}, we have the optimization problem
	\begin{equation}
		\vartheta = arg\max_{t^\prime \in \Theta^\prime} \left\{M \sum_{j=1}^M n_j  \sum_{i=1}^{n_j} \omega_{j}  \omega_{i|j} \log f(s_{ji};t^\prime)\right\},
	\end{equation}
	where $(\omega_{1},\cdots,\omega_{M})\sim Dirichlet (1,\cdots,1)$ and $(\omega_{1|j},\cdots,\omega_{n_j|j})\sim Dirichlet (1,\cdots,1)$, which are independent for all marginal and conditional probability distributions; multiplication by $M$ and $n_j$ are to match scales correctly. {This construction is similar to the weighted likelihood bootstrap approach proposed in \cite{newton1994approximate}, but the weights are updated using an enriched Dirichlet model \citep{wade2011enriched}. Further, such a construction is semi-parametric in the sense that the working model is specified in terms of a finite-dimensional parameter set, whereas the true data-generating distribution $F_O$ is represented by an infinite-dimensional parameter set and assigned to a dependent Dirichlet process prior. In addition, it is worth noting that such a specification can also accommodate nuisance models in the estimation of quantities of interest, as observed, for instance, in causal settings \citep{saarela2015Bayesian,10.1214/22-BA1322,sabbagh2026posterior}.
    Section B of the supplementary material describes the algorithm to sample from the resultant posterior distribution following the proposed approach. }


	\section{Illustrations}\label{sec:illust}
		
	{This section illustrates the performance of the proposed method under three distinct settings. Specifically, we consider: (i) hierarchical modeling with two levels of hierarchy—cluster and individual—and no covariates, where the target parameter is a population effect; (ii) hierarchical modeling with covariates, again employing two hierarchical levels and targeting a population effect; and (iii) hierarchical modeling with covariate summary, focusing on propensity score regression in scenarios of confounding adjustment. Additionally, we present results for three datasets: (i) radon data \citep{gelman2007data}, which illustrates the case of hierarchical modeling and no covariates, (ii) Programme for International Student Assessment (PISA) 2022 data \citep{OECD2024PISA}, which illustrates the case of hierarchical modeling with covariates, and (iii) tuberculosis data in Brazil \citep{nobre2023impact}, which represents an application to hierarchical modeling with covariate summary.}

	\subsection{Multilevel design}\label{sec:lmm}
	
	\subsubsection{Simulation set-up}
	
	Consider a dataset generated according to
	\begin{equation}
		\begin{split}
			Y_{ji} &= \beta_{0j} + \beta_{1j}Z_{ji} + \epsilon_{ji}, j=1,\cdots,M, i=1,\cdots,n_j\\
			Z_{ji} &= \nu_{j} + \upsilon_{ji},
		\end{split}
		\label{eq:sim_dgp01}
	\end{equation}
    where $\beta_{0j} \overset{\text{i.i.d.}}{\sim} \mathrm{Normal}(\beta_0,1)$ and $\beta_{1j} \overset{\text{i.i.d.}}{\sim} \mathrm{Normal}(\beta_1,r^2)$ are independent for all $j$. Further, we assume $\nu_{j}\sim \mathrm{Normal}(0,1)$ and $\upsilon_{ji}\sim \mathrm{Normal}(0,1)$, for all $i$ and $j$, while assuming that all pairs of indices are independent, both within and across clusters.
	
	The simulated example investigates different scenarios in the data-generating mechanism by varying the variance of the cluster level effect ($r^2$), the cluster sample size, $n_j$, and the number of cluster units $M$. Specifically, we first consider $r$, that is, the prior standard deviation for the cluster slope: varying $r \in \{0,0.2,0.4\}$. Next, we explore $n_j$, that is, cluster sizes, by assuming $n_j \in \{200, 400, 600, 800 \}$ and $ n_j = n$ for all $j$. We then analyze $M$, varying the cluster configuration by assuming that $M \in \{30,50,70\}$.  
    Finally, the regression parameters $(\beta_0,\beta_1)^\top$ are fixed at $(1,0)^\top$ in the data-generating mechanism; the target parameter is $\beta_1$.
	
	The fitted models have the general structure $Y_{ji} = \phi + \tau Z_{ji} + \epsilon_{ji}$. Results for our proposed method, the Bayesian bootstrap with enriched Dirichlet weights (EDW-BB), are presented in Table \ref{tab:resLMM3}; {results for a regular regression model (named unweighted) with flat priors, and no clustering, and regular Bayesian bootstrap (named regular BB) are presented in Tables \ref{tab:resLMM1} and \ref{tab:resLMM2} in the supplementary material}. The metrics under consideration are the average bias, the average posterior standard deviation and the coverage rates with nominal level fixed at 95\%.
	\begin{table}[!htb]
		\centering
		{\small		\caption{Summary of the estimates (bias, standard deviation, coverage) of $\tau$ over 1,000 Monte Carlo replicates and 1,000 bootstrap samples for the EDW-BB model under a multilevel data-generating process. Results for variations in $r$, $n_j$, and $M$. No covariate adjustment. 
        }
			\label{tab:resLMM3}
			\renewcommand{\arraystretch}{1}
			\begin{tabular}{ll|rrr|rrr|rrr}
				\hline
				\multicolumn{2}{c|}{\multirow{3}{*}{EDW-BB}}	& \multicolumn{3}{c|}{$r=0$} &  \multicolumn{3}{c|}{$r=.2$} & \multicolumn{3}{c}{$r=.4$} \\
				\multicolumn{2}{c|}{} & \multicolumn{3}{c|}{$M$} &  \multicolumn{3}{c|}{ $M$} & \multicolumn{3}{c}{ $M$} \\
				& &$30$ & $50$ & $70$ & $30$ & $50$ & $70$ & $30$ & $50$ & $70$\\
				\hline \hline
				\parbox[r]{4mm}{
    \multirow{4}{*}{
        \raisebox{-0.1mm}{
            \rotatebox[origin=r]{90}{Bias}
        }
    }
}		&$n_j=200$  & .001  & -.002  & -.001  & .002  & .002  & .003  & .001  & .004  & -.003\\
				&$n_j=400$  & .003  & -.004  & -.002  & .008  & ${\tiny{<}}\lvert\!.001\!\rvert$  & .002  & -.002  & -.003  & .003\\
				&$n_j=600$  & -.002  & -.002  & -.003  & .001  & .001  & .003  & .003  & .003  & -.005\\
				&$n_j=800$  & -.006  & -.001  & -.002  & ${\tiny{<}}\lvert\!.001\!\rvert$  & .001  & -.001  & -.003  & -.004  & .002\\
				\hline \hline
				\parbox[r]{4mm}{\multirow{4}{*}{\rotatebox[origin=r]{90}{{\shortstack{Standard \ \\deviation}}}}}	& $n_j=200$  & .079  & .065  & .056  & .088  & .071  & .062  & .110  & .090  & .078\\
				&$n_j=400$  & .078  & .064  & .056  & .087  & .072  & .063  & .110  & .090  & .078\\
				&$n_j=600$  & .079  & .064  & .056  & .088  & .071  & .062  & .109  & .089  & .078\\
				&$n_j=800$  & .078  & .064  & .056  & .086  & .071  & .062  & .109  & .089  & .078\\
				\hline \hline
				\parbox[r]{4mm}{
    \multirow{4}{*}{
        \raisebox{-0.1mm}{
            \rotatebox[origin=r]{90}{Coverage}
        }
    }
}&$n_j=200$ & 90.5 & 91.5 & 94.2 & 89.2 & 92{.0} & 94.2 & 90.8 & 93.3 & 90.3\\
				&$n_j=400$ & 89.9 & 92.3 & 92{.0} & 90.1 & 91.5 & 92.8 & 90.7 & 91.5 & 93{.0}\\
				&$n_j=600$ & 87.9 & 91.3 & 91.3 & 90.6 & 92.8 & 92.1 & 90.9 & 92.6 & 91{.0}\\
				&$n_j=800$ & 89.2 & 91.7 & 92.2 & 89.9 & 91.4 & 92.3 & 91.6 & 93{.0} & 92.5\\
				\hline
		\end{tabular}}
	\end{table}
	
    We observe unbiased estimates and standard deviations that converge to zero as the sample size increases across all scenarios considered. We further observe that the standard deviations converge to zero more slowly for the EDW-BB model than for the unweighted and regular BB models. Regarding coverage rates, we observe poor coverage in the unweighted and regular BB models, whereas our proposed method achieves values above 90\% across all scenarios under consideration. Complementary results are presented in Table \ref{tab:resLMM3extra_0}, considering $M \in \{100,200\}$ and $n_j \in \{1,000\,\,,\,2,000\}$. As the number of clusters and units per cluster increase, we observe coverage rates closer to the nominal level when analyzing (symmetric) credible intervals with $60\%$, $80\%$, and $95\%$ posterior probability.
	
	\begin{table}[!htb]
		\centering
		{\small	\caption{Summary of the estimates of $\tau$ over 1,000 Monte Carlo replicates and 1,000 bootstrap samples for the EDW-BB model under a multilevel data-generating process. Additional results of coverage rates for $M \in \{100,200\}$ and $n_j \in \{1,000\,\,,\,2,000\}$. No covariate adjustment. 
        }
			\label{tab:resLMM3extra_0}
			\begin{tabular}{lll|rr|rr|rr}
				\hline
				\multicolumn{3}{c|}{\multirow{3}{*}{EDW-BB}}	& \multicolumn{2}{c|}{$r=0$} &  \multicolumn{2}{c|}{$r=.2$} & \multicolumn{2}{c}{$r=.4$} \\
				\multicolumn{3}{c|}{}	& \multicolumn{2}{c|}{ $M$} &  \multicolumn{2}{c|}{$M$} & \multicolumn{2}{c}{$M$} \\
				& & &$100$ & $200$ &$100$ & $200$ &$100$ & $200$\\
				\hline \hline
				\parbox[r]{2mm}{\multirow{6}{*}{\rotatebox[origin=r]{90}{{Coverage:}}}}& \parbox[r]{2mm}{\multirow{2}{*}{\rotatebox[origin=r]{90}{{60\%}}}}		&$n_j=1,000$  &  59.8 &  56.2 & 55.7  & 57.9  & 57.4  & 59.2 \\
				&&$n_j=2,000$  &  57.3 &  60{.0} &  56.2 & 61.3 & 57.3  & 57.2 \\
				\cline{2-9} \noalign{\vskip\doublerulesep
         \vskip-\arrayrulewidth} \cline{2-9}
				&\parbox[r]{2mm}{\multirow{2}{*}{\rotatebox[origin=r]{90}{{80\%}}}}	&$n_j=1,000$  &  79.8 &  75.9 & 76.9  & 76.9  & 77.9  & 78.5 \\
				&&$n_j=2,000$  &  77.6 &  80{.0} & 76.5  &  80.9 & 78.6  & 77.3 \\
				\cline{2-9} \noalign{\vskip\doublerulesep
         \vskip-\arrayrulewidth} \cline{2-9}
				&\parbox[r]{2mm}{\multirow{2}{*}{\rotatebox[origin=r]{90}{{95\%}}}}		&$n_j=1,000$  & 94.6  & 93.7  & 93.4  & 93.4  & 93.3   & 93{.0} \\
				&&$n_j=2,000$  &  93.6 & 95{.0} & 91.2  &  94.9 & 93.7  & 94.9 \\
				\hline
		\end{tabular}}
	\end{table}
	
	\subsubsection{Real data: Radon data analysis}
	
	The proposed method is now applied to the radon data in Minnesota \citep[Chapter 12]{gelman2007data}. The analysis involves modeling the log radon activity ($y$) as a function of a binary floor measurement variable ($x$: indicating the first floor), while accounting for clustering at the county level in Minnesota. The dataset comprises $919$ observations across $85$ counties. The average number of measurements per county is $10.81$, with a minimum of $1$ and a maximum of $116$, and a standard deviation of $18.67$. Further, the frequency of first-floor measurements in the dataset is $153$, and the average log radon activity is $1.327$ for $x=0$, and $0.713$ for $x=1$.

	Table \ref{tab:radon-md} shows all models considered in the analysis. Models M5 and M6 are those that adopt a decision-theoretic approach to infer the association of interest, with M6 representing the proposed method,     
    whereas M1, M2, M3 and M4 are standard variations of multilevel models; $x^\ast$ denotes the centered version of $x$.
	\begin{table}[!htb]
		\centering
		\caption{Description of models considered for adjustment in the application for the Minnesota Radon data.}
		\label{tab:radon-md}
		\begin{tabular}{l|lll}
			\hline
			Name & Model & Random effects &  Weights\\
			\hline
			M1 & $y_{ji} = \beta_{0j} + \beta_{1}x^\ast_{ji} + \epsilon_{ji} $ & $\beta_{0j} \sim Normal (\beta_0,\varrho_0^2)$ & None\\ 
			M2 & $y_{ji} = \beta_0 + \beta_{1j}x^\ast_{ji} + \epsilon_{ji} $ & $\beta_{1j} \sim Normal (\beta_1,\varrho_1^2)$& None\\ 
			M3 & $y_{ji} = \beta_{0j} + \beta_{1j}x^\ast_{ji} + \epsilon_{ji} $ &$\beta_{lj} \overset{ind.}{\sim} Normal (\beta_{l},\varrho_l^2)$, $l=0,1$& None\\ 
            M4 & $y_{ji} = \beta_0 + \beta_{1}x^\ast_{ji} + \epsilon_{ji} $ & None & None\\ 
			M5 & $y_{ji} = \beta_0 + \beta_{1}x^\ast_{ji} + \epsilon_{ji} $ &None& BB\\ 
			M6 & $y_{ji} = \beta_0 + \beta_{1}x^\ast_{ji} + \epsilon_{ji} $ &None& EDW-BB\\ 
			\hline
		\end{tabular}
	\end{table}

\begin{figure}[!htb]
		\centering
		\includegraphics[scale=.4]{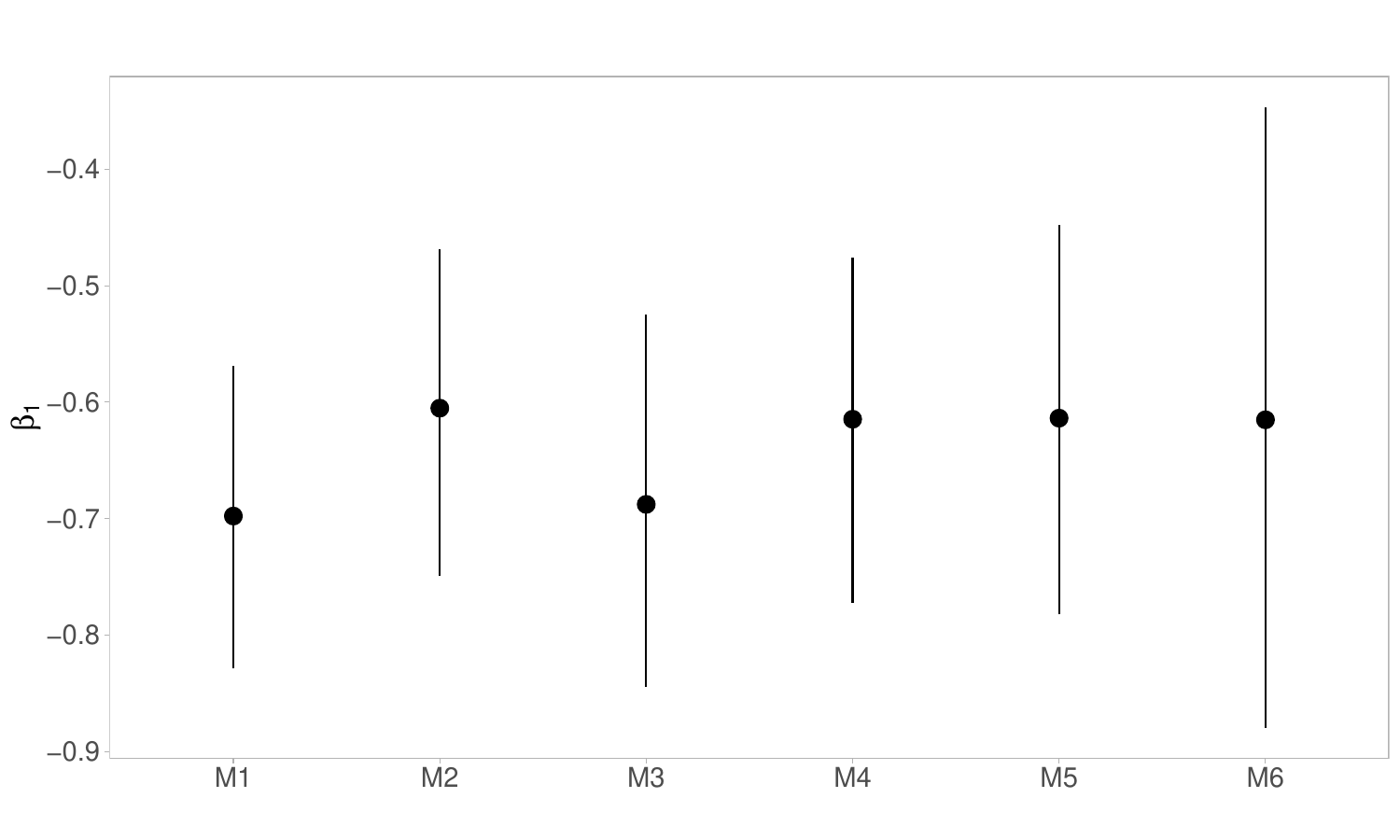}
		\caption{Posterior 95\% credible intervals (segments) for the population effect of $x$ on $y$ in the radon data analysis. Solid circles represent the posterior mean for each model.}
		\label{fig:radon-res}
	\end{figure}

    The estimand of interest is the fixed effect of $x$ on $y$. Posterior $95\%$ credible intervals of the target parameter ($\beta_1$) for each model are presented in Figure \ref{fig:radon-res}. Initially, we note that the posterior summaries are similar for M1 and M3, as well as for M2, M4 and M5. Further, models M2 and M3, which consider a slope-varying coefficient for $x$, exhibit different posterior means, with M3 displaying a slightly wider credible interval. On the other hand, M6 yields a posterior mean similar to those of M2, M4 and M5, but with a substantially wider credible interval that encompasses the credible intervals of all other models, including M1 and M3. It is important to note that the proposed method (M6) makes no distributional assumption about $F_O$, whereas the multilevel models assume the Gaussian random effects are a good fit while targeting unmeasured components under $F_O$. { Consequently, we attribute the observed differences in variability between M6 and models M1, M2 and M3 to these model assumptions; the differences between M6 and models M4 and M5 are consistent with the findings from the simulation studies presented above. Finally, based on our simulation studies (see Tables \ref{tab:resLMM3} and \ref{tab:resLMM3extra_0} and Tables \ref{tab:resLMM1} and \ref{tab:resLMM2} in the Supplementary Material for details), when a random slope for $x$ is present in the data-generating process, omitting it—as in models M4 and M5—may lead to an underestimation of the posterior variance and, consequently, to undercoverage. In such cases, M6 is expected to better reflect the underlying uncertainty than M4 and M5.} Comparing M6 with M2 and M3, we note that the observed differences in the posterior summaries may lead to different interpretations.

	{\subsection{Multilevel design with covariate adjustment}\label{sec:covadj}

\subsubsection{Simulation set-up} Let the data-generating process be defined as,
\begin{equation}
\begin{split}
Y_{ji} &= \beta_{0j} + \beta_{1j}Z_{ji} + \sum_{k=1}^{3}\beta_{Xk}X_{kji} + \sum_{k=1}^{3}\beta_{Wk}W_{kj} + \epsilon_{ji},\\
Z_{ji} &= \nu_{j} + \upsilon_{ji},  \; j=1,\cdots,M,\; i=1,\cdots,n_j,
\end{split}
\label{eq:sim_dgp02}
\end{equation}
where $\beta_{0j} \overset{\text{i.i.d.}}{\sim} \mathrm{Normal}(\beta_0,1)$ and $\beta_{1j} \overset{\text{i.i.d.}}{\sim} \mathrm{Normal}(\beta_1,r^2)$, mutually independent for all $j$. Further, we assume
\[
(X_{1ji},X_{2ji},X_{3ji})^\top \sim \mathrm{Normal}(\mu_X,\Sigma_X) \qquad (W_{1j},W_{2j},W_{3j})^\top \sim \mathrm{Normal}(\mu_W,\Sigma_W)
\]
are the individual- and cluster-level observed covariates, respectively, where $\mu_X=(-1,2,0.5)^\top$, $\mu_W=(1.5,2,-1)^\top$, $(\Sigma_X)_{ab}=0.5^{|a-b|}$, and $(\Sigma_W)_{ab}=0.8^{|a-b|}$ for $a,b=1,2,3$. We assume that $\nu_{j}\sim \mathrm{Normal}(0,1)$, $\upsilon_{ji}\sim \mathrm{Normal}(0,1)$ and $\epsilon_{ji}\sim \mathrm{Normal}(0,0.2^2)$, for all $i$ and $j$, while assuming that all pairs of indices are independent, both within and across clusters. We also fix $\beta_{X} = (1,1,1)^\top$ and $\beta_{W} = (2,2,2)^\top$. Finally, we fix  $(\beta_0,\beta_1) = (1,0)^\top$; the target parameter is $\beta_1$.

Here, we investigate the same configurations as described in Section \ref{sec:lmm}, while assuming that the fitted models have the general structure,
\[
Y_{ji} = \phi + \tau Z_{ji} + \sum_{k=1}^{3}\phi_{Xk}X_{kji} + \sum_{k=1}^{3}\phi_{Wk}W_{kj} + \epsilon_{ji}.
\]

Tables \ref{tab:resLMM_new} and \ref{tab:resLMM3extra} show the results of our proposed method (results for the unweighted and regular BB are presented in Tables \ref{tab:resCOVadjUNW} and \ref{tab:resCOVadjBB} in the supplementary material). The conclusions are similar to those in Section \ref{sec:lmm}. Specifically, the proposed method yields unbiased estimates and larger posterior standard deviations than the unweighted and regular BB methods, while coverage rates approach the nominal levels as the number of clusters and the number of units per cluster increase.

    \begin{table}[!htb]
	\centering
	{\small
		\caption{Summary of the estimates (bias, standard deviation, coverage) of $\tau$ over 1,000 Monte Carlo replicates and 1,000 bootstrap samples for the EDW-BB model under a multilevel data-generating process. Results for variations in $r$, $n_j$, and $M$. With covariate adjustment.}
		\label{tab:resLMM_new}
		\renewcommand{\arraystretch}{1}
		\begin{tabular}{ll|rrr|rrr|rrr}
			\hline
			\multicolumn{2}{c|}{\multirow{3}{*}{EDW-BB}}
			& \multicolumn{3}{c|}{$r=0$}
			& \multicolumn{3}{c|}{$r=.2$}
			& \multicolumn{3}{c}{$r=.4$} \\
			\multicolumn{2}{c|}{}
			& \multicolumn{3}{c|}{$M$}
			& \multicolumn{3}{c|}{$M$}
			& \multicolumn{3}{c}{$M$} \\
			& &$30$ & $50$ & $70$ & $30$ & $50$ & $70$ & $30$ & $50$ & $70$\\
			\hline \hline

			\parbox[r]{4mm}{
				\multirow{4}{*}{
					\raisebox{-0.1mm}{
						\rotatebox[origin=r]{90}{Bias}
					}
				}
			}
			&$n_j=200$ & -.001 & -.001 & .002 & -.004 & -.005 & -.003 & ${\tiny{<}}\lvert\!.001\!\rvert$ & .001 & .003\\
			&$n_j=400$ & .003 & .002 & -.001 & -.003 & -.001 & .002 & -.006 & .002 & .004\\
			&$n_j=600$ & -.002 & .002 & -.001 & -.001 & ${\tiny{<}}\lvert\!.001\!\rvert$ & ${\tiny{<}}\lvert\!.001\!\rvert$ & .003 & .002 & .001\\
			&$n_j=800$ & ${\tiny{<}}\lvert\!.001\!\rvert$ & .002 & -.001 & .009 & -.001 & -.005 & -.001 & ${\tiny{<}}\lvert\!.001\!\rvert$ & .005\\

			\hline \hline

			\parbox[r]{4mm}{
				\multirow{4}{*}{
					\rotatebox[origin=r]{90}{{\shortstack{Standard \\ deviation}}}
				}
			}
			&$n_j=200$ & .071 & .060 & .053 & .079 & .067 & .059 & .101 & .085 & .075\\
			&$n_j=400$ & .070 & .060 & .053 & .079 & .067 & .059 & .099 & .085 & .074\\
			&$n_j=600$ & .071 & .060 & .053 & .079 & .067 & .059 & .099 & .084 & .074\\
			&$n_j=800$ & .070 & .060 & .053 & .079 & .067 & .059 & .100 & .084 & .074\\

			\hline \hline

			\parbox[r]{4mm}{
				\multirow{4}{*}{
					\raisebox{-0.1mm}{
						\rotatebox[origin=r]{90}{Coverage}
					}
				}
			}
			&$n_j=200$ & 89.4 & 91.2 & 93.1 & 88.7 & 88.7 & 91.7 & 89.6 & 90.8 & 92.7\\
			&$n_j=400$ & 87.8 & 93.3 & 91.0 & 90.2 & 90.6 & 90.7 & 86.6 & 90.4 & 90.4\\
			&$n_j=600$ & 88.1 & 90.0 & 92.1 & 88.7 & 91.1 & 92.0 & 88.3 & 90.6 & 93.5\\
			&$n_j=800$ & 87.5 & 91.7 & 93.5 & 89.5 & 90.5 & 92.2 & 88.7 & 89.7 & 90.6\\

			\hline
		\end{tabular}}
\end{table}

    \begin{table}[!htb]
	\centering
	{\small	\caption{Summary of the estimates of $\tau$ over $1,000$ Monte Carlo replicates and $1,000$ bootstrap samples for the EDW-BB model under a multilevel data-generating process. Additional results of coverage rates for $M \in \{100,200\}$ and $n_j \in \{1,000\,\,,\,2,000\}$. Covariate adjustment.}
		\label{tab:resLMM3extra}
		\begin{tabular}{lll|rr|rr|rr}
			\hline
			\multicolumn{3}{c|}{\multirow{3}{*}{EDW-BB}}
			& \multicolumn{2}{c|}{$r=0$}
			& \multicolumn{2}{c|}{$r=.2$}
			& \multicolumn{2}{c}{$r=.4$} \\
			\multicolumn{3}{c|}{}
			& \multicolumn{2}{c|}{$M$}
			& \multicolumn{2}{c|}{$M$}
			& \multicolumn{2}{c}{$M$} \\
			&& &$100$ & $200$ &$100$ & $200$ &$100$ & $200$\\
			\hline \hline

			\parbox[r]{2mm}{\multirow{6}{*}{\rotatebox[origin=r]{90}{{Coverage:}}}}
			& \parbox[r]{2mm}{\multirow{2}{*}{\rotatebox[origin=r]{90}{{60\%}}}}
			&$n_j=1,000$ & 56.2 & 55.7 & 56.8 & 58.4 & 52.9 & 56.9 \\
			&&$n_j=2,000$ & 57.4 & 57.8 & 58.2 & 60.5 & 54.7 & 56.6 \\
			\cline{2-9} \noalign{\vskip\doublerulesep
			\vskip-\arrayrulewidth} \cline{2-9}

			&\parbox[r]{2mm}{\multirow{2}{*}{\rotatebox[origin=r]{90}{{80\%}}}}
			&$n_j=1,000$ & 73.6 & 77.6 & 77.1 & 78.7 & 73.4 & 80.2 \\
			&&$n_j=2,000$ & 76.9 & 77.1 & 78.4 & 80{.0} & 75.6 & 76.3 \\
			\cline{2-9} \noalign{\vskip\doublerulesep
			\vskip-\arrayrulewidth} \cline{2-9}

			&\parbox[r]{2mm}{\multirow{2}{*}{\rotatebox[origin=r]{90}{{95\%}}}}
			&$n_j=1,000$ & 92.4 & 93.5 & 92.8 & 93.9 & 91.7 & 93.9 \\
			&&$n_j=2,000$ & 93.1 & 93.7 & 93.7 & 95.3 & 91.2 & 92.9 \\
			\hline
	\end{tabular}}
\end{table}

	\subsubsection{Real data: PISA 2022 data analysis}
  We now discuss our proposed method in an application involving the PISA 2022 data set. The PISA is an international initiative led by the Organization for Economic Co-operation and Development (OECD) and consists of a triennial survey of 15-year-old students conducted worldwide. The survey aims to provide evidence for international educational assessments by examining both the reproducibility of learning skills and abilities and the extrapolation of knowledge to unfamiliar settings \citep{OECD2023PISA}. The public dataset is available at \url{http://www.oecd.org/pisa/}. In our illustration, we aim to investigate the relationship between reading and math proficiency. In the PISA public database, proficiencies are represented as plausible values, which are randomly drawn from their posterior distributions given item responses, background variables, and estimated model parameters \citep{OECD2024PISA}. The present analysis focuses on the first plausible values, identified in the database by the prefix \texttt{PV1}. 
   
  In addition to the target variables, the analysis includes the following individual-level covariates: an indicator for female students, an index of economic, social, and cultural status, and an indicator for the highest parental education level, up to a Bachelor's degree or equivalent. At the country level, we include gross domestic product (GDP) per capita (US\$), total government expenditure on education (as a percentage of GDP)—both obtained from the World Bank database for 2022 \citep{wdi}—and the Human Development Index (HDI) for 2022, obtained from the United Nations Development Programme \citep{hdi}. At the school level, we include indicators for private independent and private government-dependent schools, as well as school size and the student–teacher ratio.
  
   All continuous covariates are standardized in the analysis, and the cluster index of interest is the country. Following database merging, observations with missing values in any variable are excluded, yielding a final sample of 354,340 observations across 54 countries. The number of observations per country ranges from 1,942 (Belgium) to 25,385 (Spain). 
   For model adjustment, we define math proficiency as the outcome of interest and consider models similar to those described in Table \ref{tab:radon-md}, but including the covariates listed above in addition to reading proficiency. 
   
   Given the data dimensionality, we adopt a fully conjugate structure for models M1, M2, M3, and M4, with flat priors for the fixed effects and $\mathrm{Inverse\text{-}Gamma}(2,0.001)$ priors for the variance parameters. Figure \ref{fig:pisa-res} depicts the posterior summaries of $\beta_1$ under each model. Initially, we note that models assuming a fixed slope for reading proficiency yield narrow credible intervals. Further, M2 and M3 differ in their posterior summaries, as observed in the point estimates and the credible interval bounds. On the other hand, our approach yields a wider credible interval than those observed under M2 and M3, with its upper bound lying between the upper bounds of M2 and M3. As in the radon data analysis, we attribute the differences in the credible interval ranges to the model assumptions made under $F_O$. Meanwhile, it is worth noting that our approach is computationally competitive, as it can be trivially parallelized across bootstrap indexes. In this regard, given the sample size, fitting models M1, M2, and M3 without exploiting conjugate structures would be computationally demanding, making approximate methods, such as those based on variational inference or Laplace approximation, appealing, although they are not expected to provide correct uncertainty quantification.

	\begin{figure}[!htb]
		\centering
		\includegraphics[scale=.4]{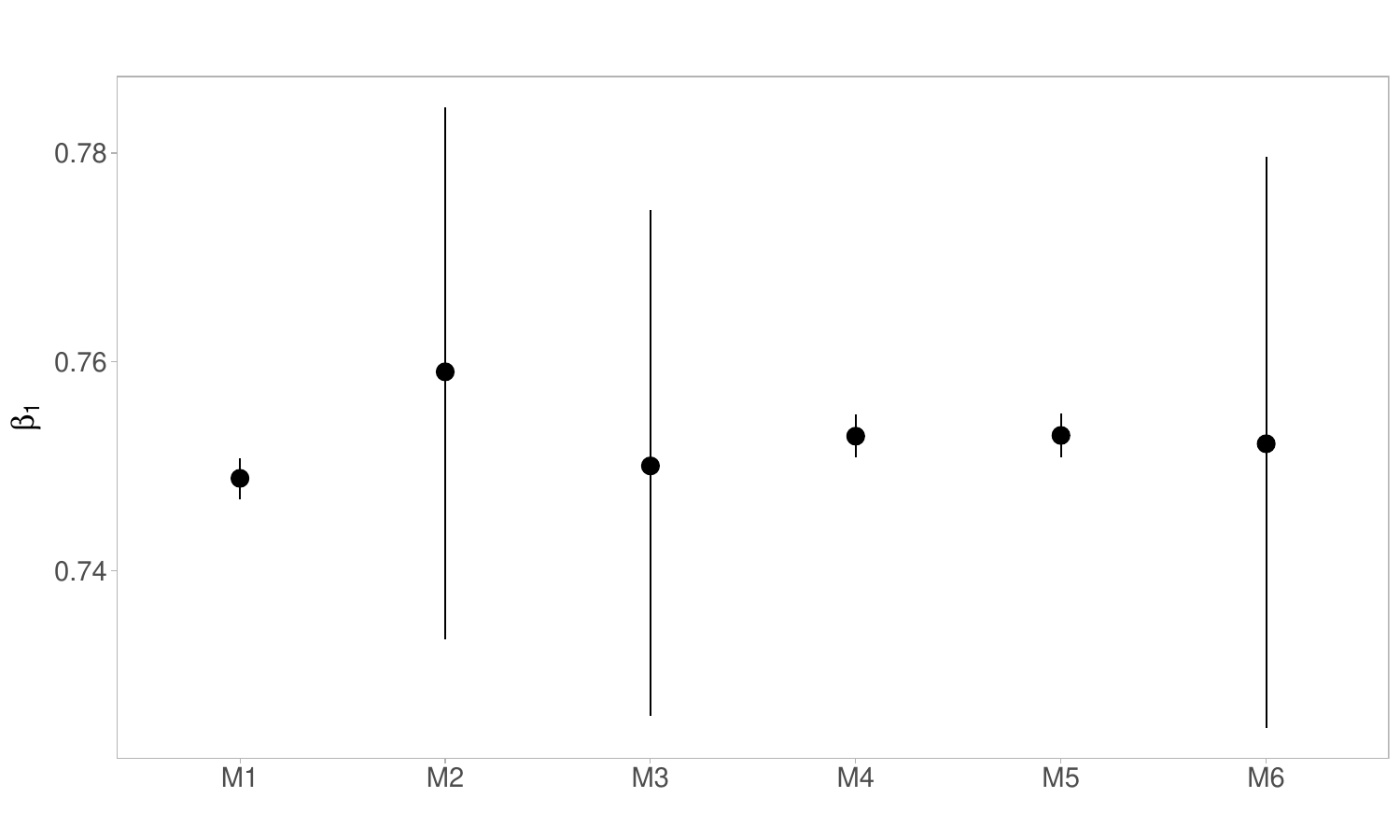}
		\caption{Posterior 95\% credible intervals (segments) for the population effect of reading score on math score in the PISA 2022 data analysis. Solid circles represent the posterior mean for each model.}
		\label{fig:pisa-res}
	\end{figure}
}
\subsection{Multilevel design with covariate summary: propensity score adjustment}\label{sec:psr}

{We now consider the setting in which the covariate set is replaced by a summary of the covariates. Specifically, we focus on propensity score adjustment with random effects to account for potential unmeasured confounding \citep{nobre2023impact}. Following \cite{10.1214/22-BA1322}, the inclusion of random effects is made through an informative prior specification, which is introduced as a penalty term in the utility model \citep{lyddon2019general,newton2021weighted} to accommodate potential unmeasured confounding in propensity score adjustment. Details of the implementation are given in Section \ref{app:psr} of the Supplementary Material. }

	\subsubsection{Simulation set-up}

    Consider the following data-generating process. First, let $U_{qj} \overset{\mathrm{i.i.d.}}{\sim} \mathrm{Normal}(0,1)$, mutually independent for $q=1,2$, and define $T_j = U_{1j}$ and $W_j = \alpha U_{1j} + (1-\alpha)U_{2j}$, where $\alpha$ controls the correlation between $T_j$ and $W_j$. Specifically, $\alpha=0$ implies independence between $T_j$ and $W_j$, whereas $\alpha=1$ implies $T_j=W_j$. The pair $(T_{j},W_{j})$ is treated as unobserved in the data analysis. Next, let $X_{kji}=\kappa_{kj}+\upsilon_{kji}$, where $\kappa_{kj} \overset{\mathrm{i.i.d.}}{\sim} \mathrm{Normal}(0,0.4^2)$ and $\upsilon_{kji} \overset{\mathrm{i.i.d.}}{\sim} \mathrm{Normal}(0,0.1^2)$, mutually independent for $k \in \{0,1\}$, be the set of observed covariates in the data analysis. 
	Finally, conditional on $(T_{j}$, $W_{j})$, $X_{1ji}$ and $X_{2ji}$, let the exposure and outcome be generated according to
	\begin{equation}
		\begin{split}
			Z_{ji} &= \gamma_0 + \gamma_1X_{1ji} + \gamma_2X_{2ji} + \rho_T T_{j} + \varepsilon_{ji},\; \varepsilon_{ji} \sim Normal (0,0.2^2),\\  
			Y_{ji} &= \tau_{j}Z_{ji} +  {X}_{ji}^\ast{\beta} + \delta_W W_{j}+ \epsilon_{ji},\; \epsilon_{ji} \sim Normal (0,0.2^2),  
		\end{split}
		\label{eq:sim_dgp1}
	\end{equation}
	where $\mathbf{X}_{ji}^\ast =(1,X_{1ji},X_{2ji},X_{1ji}X_{2ji},X_{1ji}W_{j},X_{2ji}W_{j},X_{1ji}X_{2ji}W_{j})$. The model parameters are fixed at $(\gamma_0,\gamma_1,\gamma_2)= (1,1,1)^\top$ and $\beta  = (1,1,1,1,0,0,0)^\top$. We also fix $\delta_W = \rho_T = 1$. 
	
	The simulation analyses follow by investigating variations in the strength of unmeasured confounding, the number of clusters, the number of units per cluster and the distribution of $\tau_{j}$. Initially, we consider variations in the strength of unmeasured confounding based on the following specifications of $\alpha$: (i) no unmeasured confounding: $\alpha = 0$, i.e., $Corr(T_j,W_j)=0$; (ii) moderate unmeasured confounding: $\alpha$ chosen such that the theoretical correlation between $T_j$ and $W_j$ is $0.7$; and (iii) perfect unmeasured confounding, i.e., $\alpha = 1$.
	Then, we assume that $M \in \{30,50\}$, and consider the number of units (replicates) varying in the set $\{30,40,50\}$. We also assume that $\tau_j \sim Normal (0,r^2)$, and allow $r$ varying in $\{0,0.2,0.4\}$. The target parameter is the average treatment effect, and its true value is zero according to the above specification.
	
	The fitted models have the following general structure,
	\begin{equation*}
		\label{eq:mod_adj2}
		\begin{split}
			Z_{ji} &= \gamma_0 + \gamma_1X_{1ji} + \gamma_2X_{2ji} + \nu_{j} + \varepsilon_{ji},\; \varepsilon_{ji} \sim Normal (0,\varsigma^2),\\ 
			Y_{ji} &= \tau Z_{ji} + \theta_0+ \theta_1 \hat{B}_{ji} + \epsilon_{ji},\; \epsilon_{ji} \sim Normal (0,\sigma^2),
		\end{split}
	\end{equation*}
	{where $\hat{B}_{ji}$ is a summary statistic, which is represented by the estimated propensity score in this example.} A random-effect structure is introduced only in the exposure model. Thus, we propose to consider an informative prior specification only for the exposure random effect parameter $\nu_j$, whereas noninformative priors are considered for regression parameters and hyperparameters. 
	
	Table \ref{tab:rescausal3} shows a summary of the results for our proposed method (EDW-BB). Results for the unweighted method with estimated propensity score and regular Bayesian bootstrap with linked weights in propensity score regression are presented in Tables \ref{tab:rescausal1ests} and \ref{tab:rescausal2} in the Supplementary Material. For comparison, we also present results for the regular Bayesian bootstrap and EDW-BB, when marginalizing the exposure model over $\nu_j$, in Tables \ref{tab:rescausal2ml} and \ref{tab:rescausal3ml}, respectively, and assuming a known covariance structure for the propensity score model (to simplify computational cost). 
	Again, the metrics under consideration are the average bias, the average posterior standard deviation, and the coverage rates with nominal level fixed at 95\%. All models relied on 1,000 bootstrap replicates, except for those that were unweighted.
	
	First, we observe small and similar values of bias for all scenarios and models under consideration, and bias diminishes when moving from the no unmeasured confounding case ($Corr(T,W)=0$) to the perfect unmeasured confounding case ($T=W$). In addition to an analytical result to justify these findings for the weighted likelihood approaches, the results are consistent with those of \cite{nobre2023impact}, which analyzes the exact value of the bias in comparable settings. Our observation is that when there is no unmeasured confounding, including random effects in the propensity score model may distort the balance between exposure groups and induce a bias that may be as large as the scenario of unmeasured confounding. Second, in terms of average standard deviation, EDW-BB is the largest across all combinations of  $n_j$ and $M$. Finally, when analyzing coverage rates, we observe that when the data-generating process assumes a constant $\tau$ across clusters, EDW-BB tends to be overly conservative, whereas BB tends to be close to the nominal level. When $\tau$ varies stochastically across clusters, EDW-BB exhibits coverage rates decreasing above the nominal level for all variations of the strength of the unmeasured confounding. In contrast, unweighted and standard BB methods show coverage rates that decrease below the nominal level as $n_j$ increases; this effect can be stronger depending on the strength of unmeasured confounding.  Additionally, as observed in Tables \ref{tab:rescausal2ml} and \ref{tab:rescausal3ml} in the supplementary material, models that marginalize over $\nu_j$ tend to remain biased even asymptotically in the presence of unmeasured confounding, which translates into poor coverage rates (below $10\%$ in most cases) across all settings under consideration.
        	
	Table \ref{tab:rescausal1ext} shows complementary results for model EDW-BB based on $n_j\in \{80,100,120\}$ and $M \in \{30,50\}$, and additional results of coverage rates for symmetric 60\% and 80\% credible intervals over 1,000 Monte Carlo data replicates. In the presence of an unmeasured confounder, as $M$ and $n_j$ increase, we note rates close to the nominal levels mainly when $r=0.4$. 

    	\begin{table}[!htb]
		\centering
		{\small	\caption{
        Summary of the estimates (bias, standard deviation, coverage) of $\tau$ over 1,000 Monte Carlo replicates and 1,000 bootstrap samples for the EDW-BB model under a multilevel data-generating process. Results for variations in $r$, $n_j$, and $M$. Covariate summary via propensity score adjustment. 
        }
			\label{tab:rescausal3}
            \renewcommand{\arraystretch}{1.3}	
			\begin{tabular}{llll|rrr|rrr|rrr}
				\hline
				\multicolumn{4}{c|}{\multirow{3}{*}{EDW-BB}}		& \multicolumn{3}{c|}{$Cor(T,W)=0$} &  \multicolumn{3}{c|}{$Cor(T,W)=0.7$} & \multicolumn{3}{c}{$T=W$} \\
                				\multicolumn{4}{c|}{\multirow{2}{*}{}}		& \multicolumn{3}{c|}{$n_j$} &  \multicolumn{3}{c|}{ $n_j$} & \multicolumn{3}{c}{$n_j$} \\
				&&&		& \multicolumn{1}{c}{$30$} & \multicolumn{1}{c}{$40$} & \multicolumn{1}{c|}{$50$} & \multicolumn{1}{c}{$30$} & \multicolumn{1}{c}{$40$} & \multicolumn{1}{c|}{$50$} & \multicolumn{1}{c}{$30$} & \multicolumn{1}{c}{$40$} & \multicolumn{1}{c}{$50$} \\
				\hline \hline
			\parbox[r]{2mm}{\multirow{6}{*}{\rotatebox[origin=r]{90}{{Bias}}}} &\parbox[r]{2mm}{\multirow{2}{*}{\rotatebox[origin=r]{90}{{$r=0$}}}}&\parbox[r]{2mm}{\multirow{2}{*}{\rotatebox[origin=r]{90}{{$M$}}}}&$30$ & -.008 & -.008 & -.005 & -.004 & -.004 & -.002 & ${\tiny{<}}\lvert\!.001\!\rvert$ & ${\tiny{<}}\lvert\!.001\!\rvert$ & .001\\
				&&&$50$ & -.006 & -.005 & -.004 & -.003 & -.002 & -.002 & ${\tiny{<}}\lvert\!.001\!\rvert$ & ${\tiny{<}}\lvert\!.001\!\rvert$ & ${\tiny{<}}\lvert\!.001\!\rvert$\\
				\cline{2-13}
				&\parbox[r]{2mm}{\multirow{2}{*}{\rotatebox[origin=r]{90}{{$r=.2$}}}}&\parbox[r]{2mm}{\multirow{2}{*}{\rotatebox[origin=r]{90}{{$M$}}}}&$30$ & -.005 & -.009 & -.008 & -.002 & -.004 & -.004 & .001 & ${\tiny{<}}\lvert\!.001\!\rvert$ & -.001\\
				&&&$50$ & -.006 & -.006 & -.005 & -.002 & -.003 & -.002 & .002 & .001 & .001\\
				\cline{2-13}
				&\parbox[r]{2mm}{\multirow{2}{*}{\rotatebox[origin=r]{90}{{$r=.4$}}}}&\parbox[r]{2mm}{\multirow{2}{*}{\rotatebox[origin=r]{90}{{$M$}}}}&$30$ & -.008 & -.009 & -.008 & -.003 & -.005 & -.005 & .003 & ${\tiny{<}}\lvert\!.001\!\rvert$ & -.002\\
				&&&$50$ & -.005 & -.005 & -.005 & -.001 & -.002 & -.002 & .002 & .001 & .001\\
				\hline \hline
				\parbox[r]{2mm}{\multirow{6}{*}{\rotatebox[origin=r]{90}{{Standard deviation}}}}&\parbox[r]{2mm}{\multirow{2}{*}{\rotatebox[origin=r]{90}{{$r=0$}}}}&\parbox[r]{2mm}{\multirow{2}{*}{\rotatebox[origin=r]{90}{{$M$}}}}&$30$ & .059 & .051 & .046 & .057 & .050 & .045 & .058 & .051 & .046\\
				&&&$50$ & .046 & .040 & .036 & .045 & .039 & .035 & .046 & .040 & .035\\
				\cline{2-13}
				&\parbox[r]{2mm}{\multirow{2}{*}{\rotatebox[origin=r]{90}{{$r=.2$}}}}&\parbox[r]{2mm}{\multirow{2}{*}{\rotatebox[origin=r]{90}{{$M$}}}}&$30$ & .070 & .063 & .059 & .068 & .061 & .057 & .069 & .062 & .058\\
				&&&$50$ & .055 & .049 & .046 & .053 & .048 & .045 & .054 & .049 & .046\\
				\cline{2-13}
				&\parbox[r]{2mm}{\multirow{2}{*}{\rotatebox[origin=r]{90}{{$r=.4$}}}}&\parbox[r]{2mm}{\multirow{2}{*}{\rotatebox[origin=r]{90}{{$M$}}}}&$30$ & .096 & .089 & .086 & .095 & .088 & .085 & .096 & .089 & .086\\
				&&&$50$ & .075 & .070 & .068 & .074 & .070 & .067 & .075 & .070 & .068\\
				\hline \hline
				\parbox[r]{2mm}{\multirow{6}{*}{\rotatebox[origin=r]{90}{{Coverage}}}}&\parbox[r]{2mm}{\multirow{2}{*}{\rotatebox[origin=r]{90}{{$r=0$}}}}&\parbox[r]{2mm}{\multirow{2}{*}{\rotatebox[origin=r]{90}{{$M$}}}}&$30$ & 99.7 & 99.7 & 99.8 & 99.6 & 99.4 & 99.7 & 99.7 & 99.4 & 99.8\\
				&&&$50$ & 99.5 & 100{.0} & 99.9 & 99.5 & 100{.0} & 99.9 & 99.8 & 100{.0} & 99.9\\
				\cline{2-13}
				&\parbox[r]{2mm}{\multirow{2}{*}{\rotatebox[origin=r]{90}{{$r=.2$}}}}&\parbox[r]{2mm}{\multirow{2}{*}{\rotatebox[origin=r]{90}{{$M$}}}}&$30$ & 98.7 & 98.1 & 98.4 & 98.8 & 98{.0} & 98.5 & 99.1 & 97.9 & 98.8\\
				&&&$50$ & 98.2 & 98.9 & 98.4 & 98.7 & 99.4 & 98.2 & 98.9 & 99.5 & 98.4\\
				\cline{2-13}
				&\parbox[r]{2mm}{\multirow{2}{*}{\rotatebox[origin=r]{90}{{$r=.4$}}}}&\parbox[r]{2mm}{\multirow{2}{*}{\rotatebox[origin=r]{90}{{$M$}}}}&$30$ & 96.8 & 94.9 & 96.3 & 97.2 & 95{.0} & 95.8 & 96.9 & 95.2 & 96.5\\
				&&&$50$ & 96.9 & 97.5 & 96.2 & 96.8 & 97.6 & 96.3 & 96.6 & 97.5 & 96.5\\
				\hline
		\end{tabular}}
	\end{table}

	\begin{table}[!htb]
		\centering
		{\small	\caption{
        Summary of the estimates (bias, standard deviation, coverage) of $\tau$ over 1,000 Monte Carlo replicates and 1,000 bootstrap samples for the EDW-BB model under a multilevel data-generating process. Results for variations in $r$, $n_j$, and $M$. Additional results of coverage rates for $M \in \{30,50\}$ and $n_j \in \{80,100,120\}$. Covariate summary via propensity score adjustment. 
        }
			\label{tab:rescausal1ext}
			\renewcommand{\arraystretch}{1.3}
			\begin{tabular}{llll|rrr|rrr|rrr}
					\hline
				\multicolumn{4}{c|}{\multirow{3}{*}{EDW-BB}}		& \multicolumn{3}{c|}{$Cor(T,W)=0$} &  \multicolumn{3}{c|}{$Cor(T,W)=0.7$} & \multicolumn{3}{c}{$T=W$} \\
                				\multicolumn{4}{c|}{\multirow{2}{*}{}}		& \multicolumn{3}{c|}{$n_j$} &  \multicolumn{3}{c|}{ $n_j$} & \multicolumn{3}{c}{$n_j$} \\
				&&&		& \multicolumn{1}{c}{$80$} & \multicolumn{1}{c}{$100$} & \multicolumn{1}{c|}{$120$} & \multicolumn{1}{c}{$80$} & \multicolumn{1}{c}{$100$} & \multicolumn{1}{c|}{$120$} & \multicolumn{1}{c}{$80$} & \multicolumn{1}{c}{$100$} & \multicolumn{1}{c}{$120$} \\
				\hline  \hline
				\parbox[r]{2mm}{\multirow{6}{*}{\rotatebox[origin=r]{90}{{Coverage: 60\%}}}} &\parbox[r]{2mm}{\multirow{2}{*}{\rotatebox[origin=r]{90}{{$r=0$}}}}&\parbox[r]{2mm}{\multirow{2}{*}{\rotatebox[origin=r]{90}{{$M$}}}}&$30$ & 84.7 & 81.9 & 82.4 & 84.2 & 82.1 & 81.3 & 84.1 & 82{.0} & 82.5\\
				&&&$50$ & 82.6 & 84.1 & 81.1 & 83.2 & 83.1 & 82.2 & 82.5 & 83.3 & 82.9\\
				\cline{2-13}
				&\parbox[r]{2mm}{\multirow{2}{*}{\rotatebox[origin=r]{90}{{$r=.2$}}}}&\parbox[r]{2mm}{\multirow{2}{*}{\rotatebox[origin=r]{90}{{$M$}}}}&$30$ & 69.2 & 67.2 & 68.3 & 69.8 & 66.8 & 67{.0} & 69{.0} & 67.6 & 66.7\\
				&&&$50$ & 68{.0} & 67.2 & 69.1 & 67.4 & 66.6 & 69.3 & 68{.0} & 67.1 & 68.3\\
				\cline{2-13}
				&\parbox[r]{2mm}{\multirow{2}{*}{\rotatebox[origin=r]{90}{{$r=.4$}}}}&\parbox[r]{2mm}{\multirow{2}{*}{\rotatebox[origin=r]{90}{{$M$}}}}&$30$ & 61.7 & 61.2 & 61.4 & 61.5 & 61.2 & 61.1 & 61.9 & 61.6 & 61.4\\
				&&&$50$ & 62.3 & 60.9 & 62.8 & 62.2 & 60.9 & 62.5 & 61.2 & 61.2 & 61.7\\
				\hline \hline
				\parbox[r]{2mm}{\multirow{6}{*}{\rotatebox[origin=r]{90}{{Coverage: 80\%}}}}&\parbox[r]{2mm}{\multirow{2}{*}{\rotatebox[origin=r]{90}{{$r=0$}}}}&\parbox[r]{2mm}{\multirow{2}{*}{\rotatebox[origin=r]{90}{{$M$}}}}&$30$ & 97{.0} & 95.9 & 96{.0} & 96.7 & 95.6 & 95.6 & 96.1 & 96.2 & 95.5\\
				&&&$50$ & 96.5 & 96.4 & 97.2 & 96.7 & 97{.0} & 96.9 & 96.6 & 97.4 & 96.3\\
				\cline{2-13}
				&\parbox[r]{2mm}{\multirow{2}{*}{\rotatebox[origin=r]{90}{{$r=.2$}}}}&\parbox[r]{2mm}{\multirow{2}{*}{\rotatebox[origin=r]{90}{{$M$}}}}&$30$ & 86.7 & 84.3 & 85.1 & 86.5 & 83.5 & 84.4 & 85.4 & 83.7 & 84.8\\
				&&&$50$ & 86.3 & 87.5 & 86.2 & 85.6 & 87.4 & 86.7 & 86.2 & 86.6 & 87.5\\
				\cline{2-13}
				&\parbox[r]{2mm}{\multirow{2}{*}{\rotatebox[origin=r]{90}{{$r=.4$}}}}&\parbox[r]{2mm}{\multirow{2}{*}{\rotatebox[origin=r]{90}{{$M$}}}}&$30$ & 81{.0} & 79.9 & 81.2 & 81.1 & 80{.0} & 79.6 & 80.6 & 80.3 & 80.1\\
				&&&$50$ & 80.8 & 81.9 & 84{.0} & 80.9 & 82.6 & 83.1 & 80.4 & 82.2 & 83{.0}\\
				\hline \hline
				\parbox[r]{2mm}{\multirow{6}{*}{\rotatebox[origin=r]{90}{{Coverage: 95\%}}}}&\parbox[r]{2mm}{\multirow{2}{*}{\rotatebox[origin=r]{90}{{$r=0$}}}}&\parbox[r]{2mm}{\multirow{2}{*}{\rotatebox[origin=r]{90}{{$M$}}}}&$30$ & 99.8 & 99.9 & 99.8 & 99.9 & 100{.0} & 99.7 & 99.9 & 99.8 & 99.8\\
				&&&$50$ & 99.9 & 99.7 & 100{.0} & 100{.0} & 99.8 & 100{.0} & 99.9 & 100{.0} & 99.9\\
				\cline{2-13}
				&\parbox[r]{2mm}{\multirow{2}{*}{\rotatebox[origin=r]{90}{{$r=.2$}}}}&\parbox[r]{2mm}{\multirow{2}{*}{\rotatebox[origin=r]{90}{{$M$}}}}&$30$ & 97.6 & 96.6 & 97.1 & 97.9 & 96{.0} & 96.9 & 98{.0} & 96.5 & 97{.0}\\
				&&&$50$ & 98{.0} & 98.4 & 96.8 & 98.1 & 98.1 & 97.1 & 97.9 & 98.1 & 96.7\\
				\cline{2-13}
				&\parbox[r]{2mm}{\multirow{2}{*}{\rotatebox[origin=r]{90}{{$r=.4$}}}}&\parbox[r]{2mm}{\multirow{2}{*}{\rotatebox[origin=r]{90}{{$M$}}}}&$30$ & 95.8 & 94.8 & 94.4 & 95.7 & 94.8 & 94.4 & 95.6 & 94.5 & 94.5\\
				&&&$50$ & 96{.0} & 96.5 & 94.9 & 95.7 & 96.2 & 94.4 & 95.5 & 95.9 & 93.9\\
				\hline
		\end{tabular}}
	\end{table}

	\subsubsection{Real data: Tuberculosis data analysis}

   {We revisit the example from \cite{nobre2023impact}, which mitigates unmeasured confounding issues by incorporating random effects into the propensity score and outcome models when evaluating the effect of directly observed therapy (DOT) for tuberculosis treatment.} The dataset consists of individual-level records on new tuberculosis cases with a final diagnosis made in 2016 in the state of São Paulo, Brazil. The goal is to investigate the impact of DOT on the diagnosis of cure at the end of tuberculosis treatment. These data are now used to illustrate the EDW-BB approach.
    
   In Brazil, tuberculosis treatment lasts for at least 6 months after diagnosis and is managed by the public health system, with some peculiarities related to the country's decentralized health government structure. In particular, although the combat against tuberculosis is a national initiative, the distribution of medications to the public is organized by the municipal health administration. DOT is also administered by municipal health services and represents a gold-standard strategy for treating tuberculosis promoted by the World Health Organization (WHO) since the early 1990s; DOT involves a health professional observing or assisting patients in taking medications.

   In this analysis, we consider the odds ratio (OR) of DOT ($Z$) on the diagnosis of cure at the end of tuberculosis treatment ($Y$) as the effect of interest. The cluster indices are the health regions in São Paulo; health regions are groups of cities, defined by the unified Brazilian health system, that share similar goals and policies. {This differs from \cite{nobre2023impact}, where cities were used as the cluster index. The data consist of $12,057$ observations across 45 health regions, of which $10,462$ received a diagnosis of cure at the end of tuberculosis treatment. The average number of measurements per health region is $267.9$, with a minimum of $1$ and a maximum of $4,652$, and a standard deviation of $717.5$. Additionally, the total number of applied DOTs is $6,929$, of which $6,774$ obtained a diagnosis of cure at the end of tuberculosis treatment.}

	The set of covariates ($\mathbf{X}$) include unit- and municipal-level covariates. At the unit-level, we have indicator variables for diagnosis of Acquired Immunodeficiency Syndrome (AIDS), diagnosis of diabetes, reporting (illicit) drug use, diagnosis of alcoholism, being homeless, gender, whether currently a prisoner, diagnosis of a mental illness, and current smoking status. Additionally, we have available the type of tuberculosis and age (in years) for each individual. At the municipal level, we have the Human Development Index (HDI), a proxy for a city's socio-economic environment. 
	
	Let $Z_{ji}$ be an indicator variable for whether individual $i$ in health region $j$ was assigned to DOT. Then, we assume the following specification of the propensity score model, \[
	\begin{split}
		Z_{ji}|\mathbf{X}_{ji},\nu_j,\boldsymbol{\alpha}&\sim Bern\left(\pi_{ji}\right)\\
		{\rm logit} \left(\pi_{ji}\right) & = \mathbf{X}_{ji}^\top \boldsymbol{\alpha} + \nu_j,\; \nu_j\sim \mathrm{Normal}(0,\varsigma^2),
	\end{split}
	\]
	where $\mathbf{X}_{ji}$ denotes the set of covariates for individual $i$ at health region $j$, and $\nu_j$ is a random effect placed in order to capture a latent confounding structure at health region level.
    
    Now, let $Y_{ji}$ be another indicator variable of individual $i$ in health region $j$ of diagnosis of cure at the end of tuberculosis treatment. Table \ref{tab:tb-md} summarizes the outcome models fitted, where $\hat{B}_{ji}$ denotes the estimated propensity score, used as a summary statistic for the observed covariates. 
       \begin{table}[!htb]
		\centering
		\caption{Description of models considered for adjustment in the application for tuberculosis data. Note that $\mu_{ji} = E(Y_{ji}|Z_{ji},\hat{B}_{ji})$ and $M6$ is the proposed method.}
		\label{tab:tb-md}
		\begin{tabular}{l|lll}
			\hline
			Name & Model (${\rm logit} (\mu_{ji})$) & Random effects &  Weights\\
			\hline
			M1 & $ \beta_{0j}+ \beta_{B}\hat{B}_{ji} + \beta_{1}{Z}_{ji}  $ & $\beta_{0j} \sim Normal (\beta_0,\varrho_0^2)$ & None\\ 
            M2 & $ \beta_0 + \beta_{B}\hat{B}_{ji} + \beta_{1j}{Z}_{ji}  $ & $\beta_{1j} \sim Normal (\beta_1,\varrho_1^2)$& None\\ 
			M3 & $ \beta_{0j} +\beta_B\hat{B}_{ji}+ \beta_{1j}{Z}_{ji}  $ &$\beta_{lj} \overset{ind.}{\sim} Normal (\beta_l,\varrho_l^2)$, $l=0,1$& None\\ 
			M4 & $ \beta_0 + \beta_{B}\hat{B}_{ji} + \beta_{1}{Z}_{ji} $ & None & None\\ 
			M5 & $ \beta_0 + \beta_{B}\hat{B}_{ji} + \beta_{1}{Z}_{ji} $ &None& BB\\ 
            M6 & $ \beta_0 + \beta_{B}\hat{B}_{ji}+ \beta_{1}{Z}_{ji}  $ &None& EDW-BB\\ 
			\hline
		\end{tabular}
	\end{table} 	
    Variations of multilevel models and Bayesian bootstrap are considered; M6 represents our proposed method. { From a computational perspective, model M6 required less than 2 minutes to generate 1,000 posterior samples using 26 parallel worker processes, whereas model M3 required, without accounting for propensity score adjustment, approximately 55 minutes to run four parallel chains, each of length 16,000, using the \texttt{rstanarm} package \citep{rstanarm}. All analyses were performed in R version 4.4.0 using RStudio on a same workstation equipped with an Intel Core i7-14700 processor (20 physical cores and 28 logical processors), 64 GB of RAM, and running Windows 11 Home. }

\begin{figure}[!htp]
		\centering
		\includegraphics[scale=.4]{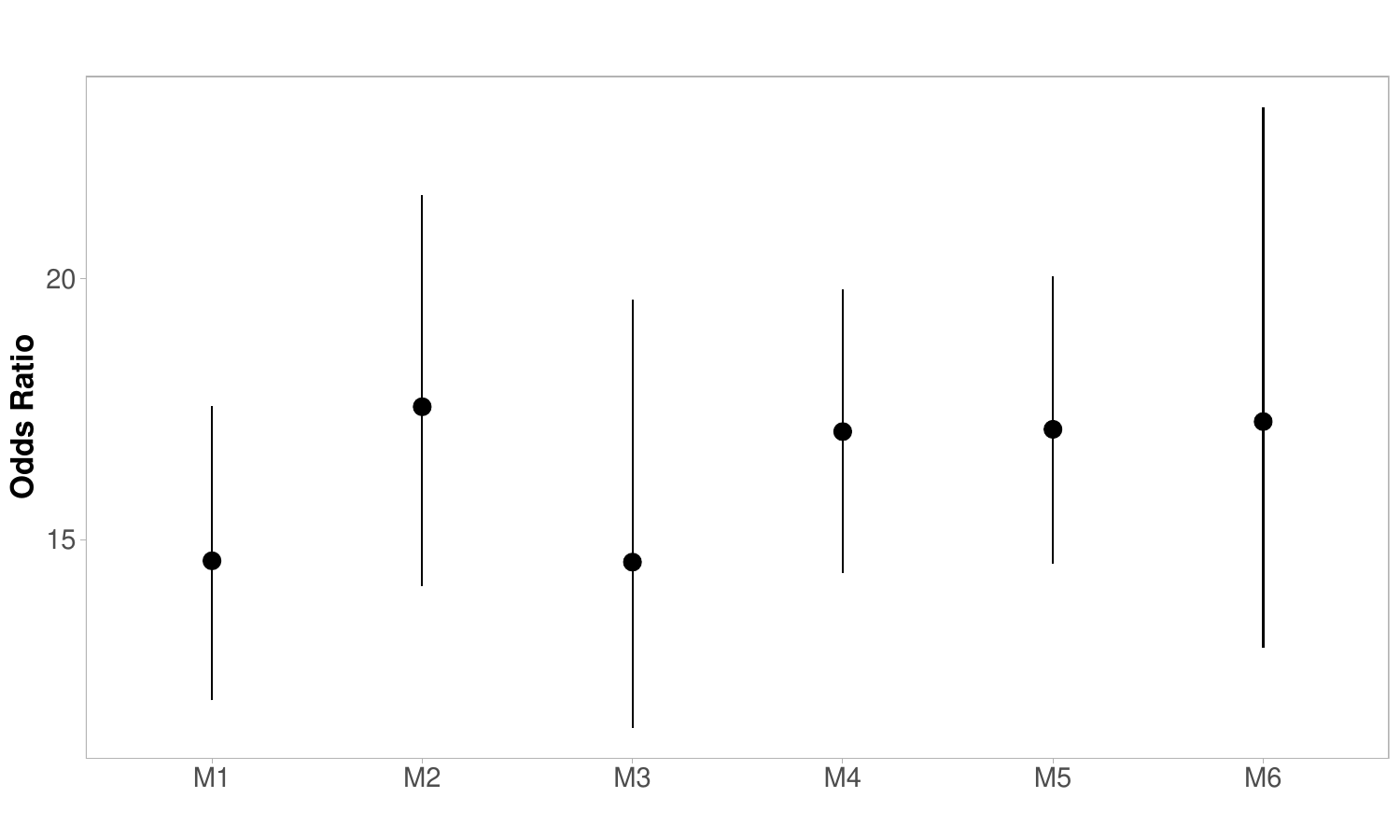}
		\caption{Posterior 95\% credible intervals for the odds ratio for DOT on the diagnosis of cure at the end of tuberculosis treatment. Circles represent the posterior mean for each model.}
		\label{fig:tb-res}
	\end{figure}

	Figure \ref{fig:tb-res} presents the posterior $95\%$ credible intervals of the odds ratio of $Z_{ji}$ for each model. Due to noncollapsibility in the proposed model, odds ratios from models with different terms in the linear predictor are not directly comparable. This distinction is important because none of the models share common predictors after incorporating propensity score estimation into the utility specification, {since different weights are being used in propensity score estimation}. The results show that M2, M4, M5, and M6 display similar posterior means, although M6 exhibits a higher posterior variance. Similarly to the discussion in the radon and PISA examples, we associate these variance behaviours to the nonparametric distributional assumptions made under $F_O$.  In contrast, models M1 and M3 yield point estimates that clearly differ from those obtained under the remaining models.

	To further clarify the above results, a plasmode simulation is provided in Section \ref{app:plamode} of the supplementary material. In this analysis, $Z$ and $X$ are treated as observed variables, and $Y$ is generated using a logistic regression model with mixed effects. Specifically, the plasmode data-generating process results from fitting a model that includes a random intercept and a random $Z$-slope across health regions, along with all covariates in the data. 
    From the results described in Table \ref{tab:tb-md-plasmode}, models M4 and M5 yield the most attractive results in terms of bias, root mean square error, and coverage rates. It should be noted, however, that previous simulation analyses have suggested that both models are sensitive to combinations of $M$, $n_j$, and the standard deviation of the slope random effect in the data-generating process (see Tables \ref{tab:rescausal1ests}, \ref{tab:rescausal2} and \ref{tab:rescausal2ext} in the Supplementary Material). The simulation also shows that the proposed method can still recover the true effect in this application, though this appears to be due to an overestimation of the posterior variance. Additionally, such an analysis highlights potential issues with the inclusion of random effects in the outcome model. Meanwhile, it is worth noting that, from the semiparametric theory \citep{tsiatis2006semiparametric}, propensity score adjustment in logistic regression does not produce an unbiased estimating function \citep{tchetgen2010doubly}, since logistic regression is not robust to mis-specification of the mean model.

	\section{Discussion}\label{sec:disc}
	
	{Hierarchical data are very common in statistical data analysis, and discussions on where and why to truncate the number of hierarchical levels in model design date back at least to \cite{lindley1972}. }
    Typically, researchers justify the inclusion of levels in the hierarchy of a particular model by performing residual analysis and following indications of non-negligible variability present in the data. In this paper, we have focused on population-level parameters as the ones that drive the model design for multilevel data, prioritizing correct inference for target parameters over model specification. In this regard, our proposed method relies on the ability of the data-generating process and model design to match partially exchangeable sequences, which are key to the estimation of target parameters. The proposal is configured as a novel Bayesian bootstrap formulation that handles multiple sources of variation and shows consistency and interesting properties as the number of clusters and the number of units per cluster increase. 
	
	The usefulness of the proposed method is illustrated through simulation studies and real-data applications. Our results suggest that the method performs well when the sequence $\psi$ is partially exchangeable. This naturally raises the question of how to diagnose such a property in practice. Recently, \cite{catalano2025measuringpartialexchangeabilityreproducing} proposed a novel metric for measuring partial exchangeability that can be evaluated from samples and may be of interest for future investigation. { Two key assumptions of our proposed method are the marginal and independence laws on the sequence $\psi$ associated with the enriched Dirichlet weights \citep{wade2011enriched}. Thus, a natural extension of this proposal is to adapt it to more general dependence settings. In particular, spatially structured observations and applications in related fields would be an interesting avenue for future investigation.}

	\FloatBarrier
		\clearpage

	\appendix
	
	\setcounter{table}{0}
	\setcounter{figure}{0}
	\setcounter{equation}{0}
	\renewcommand\thefigure{\thesection.\arabic{figure}}
	\renewcommand\thetable{\thesection.\arabic{table}}
	\renewcommand\theequation{\thesection.\arabic{equation}}

    	\section{Supplementary Material: Multilevel structure and Partial Exchangeability}\label{sec:MSOPE}
	
	Let $(\psi_{ji})_{j=1,\cdots, M;i\geq 1}$ be a random sequence with labels indicating a multilevel (or hierarchical) structure, and assume that for all finite random sequences with fixed $j$, we have
	\begin{equation}
		\left(\psi_{ji}\right)_{j;i= 1,\cdots,n_j} \overset{d}{=} \left(\psi_{j\pi(i)}\right)_{j;i= 1,\cdots,n_j},\;\text{for all}\, j,
		\label{eq:partial_exc1}
	\end{equation}
	where ${}^{``}\overset{d}{=}{}^"$ indicates equally distributed, and $\pi(\cdot)$ represents some permutation of indices $i$. Because this is valid for all $j$, it yields the following integral representation:
	\[
	p(\psi_{11},\cdots,\psi_{1n_j},\psi_{21},\cdots,\psi_{Mn_j}) = \int \prod_{j=1}^{M}\prod_{i=1}^{n_j} p\left(\psi_{ji}\mid \boldsymbol\phi_j \right)d{\Pi}(\boldsymbol\phi_1,\cdots,\boldsymbol\phi_M),
	\]
	with $(\boldsymbol\phi_1,\cdots,\boldsymbol\phi_M)$ representing a vector of latent components. Now, we assume that $(\phi_j)_{j=1,\cdots,M}$ is exchangeable itself, i.e.,
	\begin{equation}
		\left(\phi_{j}\right)_{j= 1,\cdots,M} \overset{d}{=} \left(\phi_{\pi(j)}\right)_{j= 1,\cdots,M},
		\label{eq:partial_exc2}
	\end{equation}
	in what is usually motivated by the idea that cluster parameters should not differ \textit{a priori}  \citep{bernardo1996concept}. As a result, it follows that
	\[
	{p}(\boldsymbol\phi_1,\cdots,\boldsymbol\phi_M) = \int \prod_{j=1}^{M} p\left(\phi_{j}\mid \boldsymbol\theta \right)d{Q}(\boldsymbol\theta),
	\]
	with $\boldsymbol\theta$ denoting a set of hyperparameters. Random sequences obeying \eqref{eq:partial_exc1} and \eqref{eq:partial_exc2} are said to be partially exchangeable sequences with multilevel structure.

	\clearpage

    \section{Supplementary Material: Algorithm of the proposed method}\label{sup:alg}

    \begin{algorithmic}[1]
\FOR{$l = 1$ to $B$}
    \STATE Draw random weights $(\omega_{1}^{(l)},\cdots,\omega_{M}^{(l)}) \sim Dirichlet(1,\cdots,1)$
    
    \FOR{$j = 1$ to $M$}
        \STATE Draw random weights $(\omega_{1|j}^{(l)},\cdots,\omega_{n_j|j}^{(l)}) \sim Dirichlet(1,\cdots,1)$
    \ENDFOR

    \STATE Compute
    \[
        \theta^{(l)} = \arg\max_{\theta^\prime \in \Theta^\prime} 
        \left\{ M \sum_{j=1}^M n_j \sum_{i=1}^{n_j} 
        \omega_{j}^{(l)} \omega_{i|j}^{(l)} \log f(s_{ji};\theta^\prime) \right\}
    \]
\ENDFOR

\STATE Output $\theta^{(1)}, \dots, \theta^{(B)}$.
\end{algorithmic}

	\clearpage

	\section{Supplementary Material: Additional results for the simulation data analysis presented in \ref{sec:lmm}}
	
	\begin{table}[!htb]
		\centering
		{\small	\caption{
        Summary of the estimates (bias, standard deviation, coverage) of $\tau$ over 1,000 Monte Carlo replicates and 1,000 bootstrap samples for the unweighted model under a multilevel data-generating process. Results for variations in $r$, $n_j$, and $M$.  No covariate adjustment.
        }
			\label{tab:resLMM1}
			\begin{tabular}{ll|rrr|rrr|rrr}
				\hline
                \multicolumn{2}{c|}{\multirow{3}{*}{Unweighted}}	& \multicolumn{3}{c|}{$r=0$} &  \multicolumn{3}{c|}{$r=.2$} & \multicolumn{3}{c}{$r=.4$} \\
				\multicolumn{2}{c|}{} & \multicolumn{3}{c|}{$M$} &  \multicolumn{3}{c|}{ $M$} & \multicolumn{3}{c}{ $M$} \\
				& & \multicolumn{1}{c}{$30$} & \multicolumn{1}{c}{$50$} & \multicolumn{1}{c|}{$70$} & \multicolumn{1}{c}{$30$} & \multicolumn{1}{c}{$50$} & \multicolumn{1}{c|}{$70$} & \multicolumn{1}{c}{$30$} & \multicolumn{1}{c}{$50$} & \multicolumn{1}{c}{$70$} \\
				\hline \hline
				\parbox[r]{4mm}{
    \multirow{4}{*}{
        \raisebox{-0.1mm}{
            \rotatebox[origin=r]{90}{Bias}
        }
    }
}	& 	$n_j=200$ & .001 & -.002 & -.001 & .002 & .002 & .003 & .001 & .004 & -.003\\
				& 	$n_j=400$ & .003 & -.004 & -.002 & .008 & ${\tiny{<}}\lvert\!.001\!\rvert$ & .002 & -.002 & -.003 & .003\\
				& 	$n_j=600$ & -.002 & -.002 & -.003 & .001 & .001 & .003 & .004 & .004 & -.005\\
				& 	$n_j=800$ & -.007 & -.001 & -.002 & ${\tiny{<}}\lvert\!.001\!\rvert$ & .001 & -.001 & -.003 & -.004 & .002\\
				\hline				\hline
				\parbox[r]{4mm}{\multirow{4}{*}{\rotatebox[origin=r]{90}{{\shortstack{Standard \ \\deviation}}}}} & 	$n_j=200$ & .009 & .007 & .006 & .009 & .007 & .006 & .010 & .008 & .007\\
				& 	$n_j=400$ & .006 & .005 & .004 & .007 & .005 & .004 & .007 & .006 & .005\\
				& 	$n_j=600$ & .005 & .004 & .004 & .005 & .004 & .004 & .006 & .005 & .004\\
				& 	$n_j=800$ & .005 & .004 & .003 & .005 & .004 & .003 & .005 & .004 & .003\\
				\hline \hline
				\parbox[r]{4mm}{
    \multirow{4}{*}{
        \raisebox{-0.1mm}{
            \rotatebox[origin=r]{90}{Coverage}
        }
    }
}&	$n_j=200$ & 14.2 & 15.3 & 14.8 & 14.3 & 11.8 & 15.0 & 12.9 & 12.1 & 12.5\\
				&	$n_j=400$ & 9.4 & 11.1 & 12.0 & 9.0 & 10.3 & 10.7 & 9.6 & 10.0 & 9.4\\
				&	$n_j=600$ & 9.2 & 8.9 & 7.0 & 9.2 & 8.7 & 8.7 & 9.1 & 9.4 & 6.3\\
				&	$n_j=800$ & 7.9 & 8.2 & 7.8 & 7.1 & 8.2 & 6.8 & 7.4 & 6.2 & 6.3\\
				\hline
		\end{tabular}}
	\end{table}
	
	\begin{table}[!htb]
		\centering
		{\small	\caption{
        Summary of the estimates (bias, standard deviation, coverage) of $\tau$ over 1,000 Monte Carlo replicates and 1,000 bootstrap samples for the regular Bayesian bootstrap model under a multilevel data-generating process. Results for variations in $r$, $n_j$, and $M$.  No covariate adjustment.
        }
			\label{tab:resLMM2}
			\begin{tabular}{ll|rrr|rrr|rrr}
				\hline 
                \multicolumn{2}{c|}{\multirow{3}{*}{Regular BB}}	& \multicolumn{3}{c|}{$r=0$} &  \multicolumn{3}{c|}{$r=.2$} & \multicolumn{3}{c}{$r=.4$} \\
				\multicolumn{2}{c|}{} & \multicolumn{3}{c|}{$M$} &  \multicolumn{3}{c|}{ $M$} & \multicolumn{3}{c}{ $M$} \\
				& & \multicolumn{1}{c}{$30$} & \multicolumn{1}{c}{$50$} & \multicolumn{1}{c|}{$70$} & \multicolumn{1}{c}{$30$} & \multicolumn{1}{c}{$50$} & \multicolumn{1}{c|}{$70$} & \multicolumn{1}{c}{$30$} & \multicolumn{1}{c}{$50$} & \multicolumn{1}{c}{$70$} \\
				\hline \hline
				\parbox[r]{4mm}{
    \multirow{4}{*}{
        \raisebox{-0.1mm}{
            \rotatebox[origin=r]{90}{Bias}
        }
    }
}	& $n_j=200$ & .001 & -.003 & -.001 & .002 & .002 & .003 & .001 & .004 & -.003\\
				&$n_j=400$ & .003 & -.004 & -.002 & .008 & ${\tiny{<}}\lvert\!.001\!\rvert$ & .002 & -.002 & -.003 & .002\\
				&$n_j=600$ & -.002 & -.002 & -.003 & .001 & .001 & .003 & .004 & .004 & -.005\\
				&$n_j=800$ & -.007 & -.001 & -.002 & ${\tiny{<}}\lvert\!.001\!\rvert$ & .001 & -.001 & -.003 & -.004 & .002\\
				\hline \hline
				\parbox[r]{4mm}{\multirow{4}{*}{\rotatebox[origin=r]{90}{{\shortstack{Standard \ \\deviation}}}}}&$n_j=200$ & .009 & .007 & .006 & .010 & .008 & .007 & .012 & .010 & .008\\
				&$n_j=400$ & .006 & .005 & .004 & .007 & .006 & .005 & .009 & .007 & .006\\
				&$n_j=600$ & .005 & .004 & .003 & .006 & .004 & .004 & .007 & .006 & .005\\
				&$n_j=800$ & .004 & .004 & .003 & .005 & .004 & .003 & .006 & .005 & .004\\
				\hline \hline
				\parbox[r]{4mm}{
    \multirow{4}{*}{
        \raisebox{-0.1mm}{
            \rotatebox[origin=r]{90}{Coverage}
        }
    }
}&$n_j=200$ & 14.2 & 15.7 & 15.0 & 15.3 & 13.4 & 15.9 & 15.2 & 14.6 & 15.3\\
				&$n_j=400$ & 9.3 & 11.4 & 11.7 & 9.2 & 11.0 & 11.7 & 10.5 & 12.4 & 10.9\\
				&$n_j=600$ & 9.3 & 8.8 & 7.0 & 10.3 & 9.0 & 9.2 & 10.8 & 11.0 & 7.8\\
				&$n_j=800$ & 8.1 & 8.3 & 7.9 & 7.4 & 8.6 & 7.3 & 8.4 & 6.8 & 8.4\\
				\hline
		\end{tabular}}
	\end{table}
	
	\clearpage

	\section{Supplementary Material: Additional results for the simulation data analysis presented in \ref{sec:covadj}}

    \begin{table}[!htb]
	\centering
	{\small
		\caption{
        Summary of the estimates (bias, standard deviation, coverage) of $\tau$ over 1,000 Monte Carlo replicates and 1,000 bootstrap samples for the unweighted model under a multilevel data-generating process. Results for variations in $r$, $n_j$, and $M$. With covariate adjustment.
        }
		\label{tab:resCOVadjUNW}
		\renewcommand{\arraystretch}{1}
		\begin{tabular}{ll|rrr|rrr|rrr}
			\hline
			\multicolumn{2}{c|}{\multirow{3}{*}{Unweighted}}
			& \multicolumn{3}{c|}{$r=0$}
			& \multicolumn{3}{c|}{$r=.2$}
			& \multicolumn{3}{c}{$r=.4$} \\
			\multicolumn{2}{c|}{}
			& \multicolumn{3}{c|}{$M$}
			& \multicolumn{3}{c|}{$M$}
			& \multicolumn{3}{c}{$M$} \\
			&&$30$ & $50$ & $70$ & $30$ & $50$ & $70$ & $30$ & $50$ & $70$\\
			\hline \hline

			\parbox[r]{4mm}{
				\multirow{4}{*}{
					\raisebox{-0.1mm}{
						\rotatebox[origin=r]{90}{Bias}
					}
				}
			}
			&$n_j=200$ & -.002 & -.001 & .002 & -.005 & -.005 & -.003 & ${\tiny{<}}\lvert\!.001\!\rvert$ & .001 & .003\\
			&$n_j=400$ & .003 & .002 & -.001 & -.003 & -.002 & .001 & -.006 & .002 & .003\\
			&$n_j=600$ & -.002 & .002 & -.001 & -.002 & -.001 & ${\tiny{<}}\lvert\!.001\!\rvert$ & .003 & .002 & .001\\
			&$n_j=800$ & ${\tiny{<}}\lvert\!.001\!\rvert$ & .002 & -.001 & .009 & -.001 & -.005 & -.002 & ${\tiny{<}}\lvert\!.001\!\rvert$ & .005\\

			\hline \hline

			\parbox[r]{4mm}{
				\multirow{4}{*}{
					\rotatebox[origin=r]{90}{{\shortstack{Standard \\ deviation}}}
				}
			}
			&$n_j=200$ & .009 & .007 & .006 & .009 & .007 & .006 & .010 & .008 & .007\\
			&$n_j=400$ & .006 & .005 & .004 & .006 & .005 & .005 & .007 & .006 & .005\\
			&$n_j=600$ & .005 & .004 & .003 & .005 & .004 & .004 & .006 & .005 & .004\\
			&$n_j=800$ & .005 & .003 & .003 & .005 & .004 & .003 & .005 & .004 & .004\\

			\hline \hline

			\parbox[r]{4mm}{
				\multirow{4}{*}{
					\raisebox{-0.1mm}{
						\rotatebox[origin=r]{90}{Coverage}
					}
				}
			}
			&$n_j=200$ & 16.7 & 13.6 & 14.1 & 13.9 & 14.3 & 14.4 & 11.6 & 14.0 & 11.5\\
			&$n_j=400$ & 10.9 & 12.2 & 11.4 & 10.5 & 9.2 & 11.9 & 7.5 & 8.1 & 9.8\\
			&$n_j=600$ & 7.5 & 9.3 & 7.1 & 8.3 & 6.9 & 7.3 & 5.9 & 8.0 & 6.6\\
			&$n_j=800$ & 8.3 & 7.8 & 8.3 & 6.3 & 7.0 & 7.4 & 6.2 & 5.6 & 6.7\\

			\hline
		\end{tabular}}
\end{table}

\begin{table}[!htb]
	\centering
	{\small
		\caption{
        Summary of the estimates (bias, standard deviation, coverage) of $\tau$ over 1,000 Monte Carlo replicates and 1,000 bootstrap samples for the regular Bayesian bootstrap model under a multilevel data-generating process. Results for variations in $r$, $n_j$, and $M$. With covariate adjustment.
        }
		\label{tab:resCOVadjBB}
		\renewcommand{\arraystretch}{1}
		\begin{tabular}{ll|rrr|rrr|rrr}
			\hline
			\multicolumn{2}{c|}{\multirow{3}{*}{Regular BB}}
			& \multicolumn{3}{c|}{$r=0$}
			& \multicolumn{3}{c|}{$r=.2$}
			& \multicolumn{3}{c}{$r=.4$} \\
			\multicolumn{2}{c|}{}
			& \multicolumn{3}{c|}{$M$}
			& \multicolumn{3}{c|}{$M$}
			& \multicolumn{3}{c}{$M$} \\
			&&$30$ & $50$ & $70$ & $30$ & $50$ & $70$ & $30$ & $50$ & $70$\\
			\hline \hline

			\parbox[r]{4mm}{
				\multirow{4}{*}{
					\raisebox{-0.1mm}{
						\rotatebox[origin=r]{90}{Bias}
					}
				}
			}
			&$n_j=200$ & -.002 & -.002 & .002 & -.004 & -.006 & -.003 & ${\tiny{<}}\lvert\!.001\!\rvert$ & .001 & .003\\
			&$n_j=400$ & .003 & .002 & -.001 & -.003 & -.001 & .002 & -.006 & .002 & .004\\
			&$n_j=600$ & -.002 & .002 & -.001 & -.001 & ${\tiny{<}}\lvert\!.001\!\rvert$ & ${\tiny{<}}\lvert\!.001\!\rvert$ & .003 & .002 & .001\\
			&$n_j=800$ & ${\tiny{<}}\lvert\!.001\!\rvert$ & .002 & -.001 & .010 & -.001 & -.005 & -.002 & ${\tiny{<}}\lvert\!.001\!\rvert$ & .006\\

			\hline \hline

			\parbox[r]{4mm}{
				\multirow{4}{*}{
					\rotatebox[origin=r]{90}{{\shortstack{Standard \\ deviation}}}
				}
			}
			&$n_j=200$ & .009 & .007 & .006 & .010 & .008 & .007 & .012 & .010 & .008\\
			&$n_j=400$ & .006 & .005 & .004 & .007 & .005 & .005 & .008 & .007 & .006\\
			&$n_j=600$ & .005 & .004 & .003 & .006 & .004 & .004 & .007 & .005 & .005\\
			&$n_j=800$ & .004 & .003 & .003 & .005 & .004 & .003 & .006 & .005 & .004\\

			\hline \hline

			\parbox[r]{4mm}{
				\multirow{4}{*}{
					\raisebox{-0.1mm}{
						\rotatebox[origin=r]{90}{Coverage}
					}
				}
			}
			&$n_j=200$ & 16.8 & 13.5 & 14.1 & 14.0 & 15.9 & 15.3 & 13.0 & 16.9 & 13.8\\
			&$n_j=400$ & 11.1 & 12.0 & 11.0 & 11.5 & 9.8 & 12.1 & 9.1 & 9.2 & 10.8\\
			&$n_j=600$ & 6.8 & 8.7 & 7.0 & 8.5 & 6.9 & 7.7 & 6.6 & 9.2 & 7.6\\
			&$n_j=800$ & 7.7 & 7.6 & 8.0 & 6.3 & 7.1 & 7.6 & 7.0 & 6.8 & 7.6\\

			\hline
		\end{tabular}}
\end{table}

	\clearpage

    	\section{Supplementary Material: Multilevel models with propsensity score adjustment}\label{app:psr}

        	The construction of Section \ref{sec:bbhs} presumes that the sequence of $\psi$’s is independent when conditioned on a probability measure that is partially exchangeable. In a causal inference setting, such methodology has other implications, particularly when dealing with unmeasured confounding. Specifically, it is worth noting that the plug-in method proposed in \cite{10.1214/22-BA1322} accommodates exposure and outcome models in a decision-theoretical formulation, and permits including prior information as penalty terms into the utility model \citep{lyddon2019general,newton2021weighted}. Here, we consider informative priors to account for random effects and address the latent ignorability assumption.
	
	To demonstrate how our proposal is implemented in such a causal context, let $f_{Y|Z,B}(y|z,b;\theta)$ and  $f_{Z|X}(z|x;\alpha,\nu)$ be the proposed models for the outcome and exposure units, respectively. 
	Then, the formulation from Section \ref{sec:bbhs} is adapted to the following optimization problem:
	\begin{equation}
		\arg \max_{\theta } \left\{ M\sum_{j=1}^{M}n_j\sum_{i = 1}^{n_j} \omega_{j} \omega_{i\mid j} \log f_{Y|Z,B}\left(y_{ji}|z_{ji},\widehat{b}_{ji};\theta\right) \right\},
		\label{eq:BBsample1}
	\end{equation}
	where $\widehat{b}_{ji}$ is the solution of
	\begin{equation}
		\begin{split}
			\arg \max_{\alpha,\nu} \left\{ M\sum_{j=1}^{M}n_j\sum_{i = 1}^{n_j} \omega_{j}\omega_{i\mid j} \log f_{Z|X}\left(z_{ji}|x_{ji};\alpha,\nu_j\right) + M\sum_{j=1}^{M}\tilde\omega_j \log\pi_{0}(\nu_j)\right\},
		\end{split}
		\label{eq:BBsample2}
	\end{equation}
	with $\pi_{0}$ a prior penalty term for cluster-level parameters associated with the exposure model; $\nu_{j}$ represents a random effect in the exposure model to achieve latent ignorability.
	Additionally, we have $\omega_{j}$ and $\omega_{i\mid j}$ linked weights \citep{sabbagh2026posterior}. Finally, the solutions to \eqref{eq:BBsample1} and \eqref{eq:BBsample2}, under a particular set of weights, allow for a single posterior sample of the target parameter.
	
	The above specification has two important variations. First, one may wonder about prior information being incorporated into outcome and exposure models simultaneously. Here, we focus on the case in which random effects are introduced only in the exposure model to address latent ignorability. Second, we consider the specification of $\tilde\omega_j$, $j=1,\cdots,M$, and set $\tilde{\omega}_j = 1/M$ for all $j$. Although one could instead assume prior weights drawn from a Dirichlet distribution, we view the prior specification as designed to match the conditional structure induced by Dirichlet weights, without an additional role in uncertainty quantification.

	\section{Supplementary Material: Additional results for the simulation data analysis presented in \ref{sec:psr}}

	\begin{table}[!htb]
		\centering
		{\footnotesize	\caption{Summary of the estimates (bias, standard deviation, coverage) of $\tau$ over 1,000 Monte Carlo replicates and 1,000 bootstrap samples for the unweighted model under a multilevel data-generating process. Results for variations in $r$, $n_j$, and $M$. Covariate summary via propensity score adjustment.
        }
			\label{tab:rescausal1ests}
			\renewcommand{\arraystretch}{1.4}
            			\begin{tabular}{llll|rrr|rrr|rrr}
				\hline
				\multicolumn{4}{c|}{\multirow{3}{*}{Unw+PS}}		& \multicolumn{3}{c|}{$Cor(T,W)=0$} &  \multicolumn{3}{c|}{$Cor(T,W)=0.7$} & \multicolumn{3}{c}{$T=W$} \\
                				\multicolumn{4}{c|}{\multirow{2}{*}{}}		& \multicolumn{3}{c|}{$n_j$} &  \multicolumn{3}{c|}{ $n_j$} & \multicolumn{3}{c}{$n_j$} \\
				&&&		& \multicolumn{1}{c}{$30$} & \multicolumn{1}{c}{$40$} & \multicolumn{1}{c|}{$50$} & \multicolumn{1}{c}{$30$} & \multicolumn{1}{c}{$40$} & \multicolumn{1}{c|}{$50$} & \multicolumn{1}{c}{$30$} & \multicolumn{1}{c}{$40$} & \multicolumn{1}{c}{$50$} \\
				\hline \hline
				\parbox[r]{2mm}{\multirow{6}{*}{\rotatebox[origin=r]{90}{{Bias}}}} &\parbox[r]{2mm}{\multirow{2}{*}{\rotatebox[origin=r]{90}{\small{$r=0$}}}}&\parbox[r]{2mm}{\multirow{2}{*}{\rotatebox[origin=r]{90}{{$M$}}}}&$30$ & -.005 & -.005 & .002 & -.003 & -.002 & .001 & ${\tiny{<}}\lvert\!.001\!\rvert$ & .001 & .003\\
				&&&$50$ & -.003 & -.008 & -.005 & ${\tiny{<}}\lvert\!.001\!\rvert$ & -.004 & -.002 & .004 & -.001 & .000\\
				\cline{2-13}
				&\parbox[r]{2mm}{\multirow{2}{*}{\rotatebox[origin=r]{90}{\small{$r=.2$}}}}&\parbox[r]{2mm}{\multirow{2}{*}{\rotatebox[origin=r]{90}{{$M$}}}}&$30$ & -.003 & -.005 & -.001 & ${\tiny{<}}\lvert\!.001\!\rvert$ & -.002 & ${\tiny{<}}\lvert\!.001\!\rvert$ & .003 & .001 & .001\\
				&&&$50$ & ${\tiny{<}}\lvert\!.001\!\rvert$ & -.009 & -.006 & .001 & -.005 & -.003 & .004 & -.001 & .000\\
				\cline{2-13}
				&\parbox[r]{2mm}{\multirow{2}{*}{\rotatebox[origin=r]{90}{\small{$r=.4$}}}}&\parbox[r]{2mm}{\multirow{2}{*}{\rotatebox[origin=r]{90}{{$M$}}}}&$30$ & .003 & -.006 & -.001 & .004 & -.003 & -.001 & .005 & .001 & .002\\
				&&&$50$ & .003 & -.009 & -.005 & .004 & -.004 & -.002 & .006 & -.001 & .001\\
				\hline \hline
				\parbox[r]{2mm}{\multirow{6}{*}{\rotatebox[origin=r]{90}{{Standard  deviation}}}}&\parbox[r]{2mm}{\multirow{2}{*}{\rotatebox[origin=r]{90}{\small{$r=0$}}}}&\parbox[r]{2mm}{\multirow{2}{*}{\rotatebox[origin=r]{90}{{$M$}}}}&$30$ & .197 & .169 & .147 & .108 & .095 & .080 & .063 & .060 & .046\\
				&&&$50$ & .156 & .118 & .105 & .085 & .064 & .057 & .051 & .034 & .032\\
				\cline{2-13}
				&\parbox[r]{2mm}{\multirow{2}{*}{\rotatebox[origin=r]{90}{\small{$r=.2$}}}}&\parbox[r]{2mm}{\multirow{2}{*}{\rotatebox[origin=r]{90}{{$M$}}}}&$30$ & .191 & .173 & .150 & .112 & .102 & .087 & .074 & .071 & .058\\
				&&&$50$ & .140 & .129 & .116 & .083 & .075 & .068 & .057 & .051 & .047\\
				\cline{2-13}
				&\parbox[r]{2mm}{\multirow{2}{*}{\rotatebox[origin=r]{90}{\small{$r=.4$}}}}&\parbox[r]{2mm}{\multirow{2}{*}{\rotatebox[origin=r]{90}{{$M$}}}}&$30$ & .209 & .188 & .163 & .140 & .125 & .108 & .111 & .101 & .086\\
				&&&$50$ & .163 & .141 & .127 & .109 & .094 & .084 & .087 & .076 & .068\\
				\hline \hline
				\parbox[r]{2mm}{\multirow{6}{*}{\rotatebox[origin=r]{90}{{Coverage}}}}&\parbox[r]{2mm}{\multirow{2}{*}{\rotatebox[origin=r]{90}{\small{$r=0$}}}}&\parbox[r]{2mm}{\multirow{2}{*}{\rotatebox[origin=r]{90}{{$M$}}}}&$30$ & 100.0 & 100.0 & 100.0 & 100.0 & 100.0 & 100.0 & 99.5 & 99.7 & 99.5\\
				&&&$50$ & 100.0 & 100.0 & 100.0 & 100.0 & 100.0 & 100.0 & 99.8 & 99.5 & 99.5\\
				\cline{2-13}
				&\parbox[r]{2mm}{\multirow{2}{*}{\rotatebox[origin=r]{90}{\small{$r=.2$}}}}&\parbox[r]{2mm}{\multirow{2}{*}{\rotatebox[origin=r]{90}{{$M$}}}}&$30$ & 100.0 & 100.0 & 100.0 & 100.0 & 100.0 & 99.7 & 98.9 & 99.2 & 98.0\\
				&&&$50$ & 100.0 & 100.0 & 100.0 & 99.9 & 100.0 & 99.9 & 99.0 & 98.6 & 98.2\\
				\cline{2-13}
				&\parbox[r]{2mm}{\multirow{2}{*}{\rotatebox[origin=r]{90}{\small{$r=.4$}}}}&\parbox[r]{2mm}{\multirow{2}{*}{\rotatebox[origin=r]{90}{{$M$}}}}&$30$ & 100.0 & 100.0 & 99.9 & 99.7 & 99.2 & 98.8 & 97.6 & 96.6 & 95.1\\
				&&&$50$ & 100.0 & 100.0 & 100.0 & 99.8 & 99.8 & 98.7 & 98.0 & 97.6 & 96.3\\
				\hline
		\end{tabular}}
	\end{table}

	\begin{table}[!htb]
		\centering
		{\small	\caption{Summary of the estimates (bias, standard deviation, coverage) of $\tau$ over 1,000 Monte Carlo replicates and 1,000 bootstrap samples for the regular Bayesian bootstrap model under a multilevel data-generating process. Results for variations in $r$, $n_j$, and $M$. Covariate summary via propensity score adjustment.
        }
			\label{tab:rescausal2}
			\renewcommand{\arraystretch}{1.3}
                        			\begin{tabular}{llll|rrr|rrr|rrr}
				\hline
				\multicolumn{4}{c|}{\multirow{3}{*}{Regular BB}}		& \multicolumn{3}{c|}{$Cor(T,W)=0$} &  \multicolumn{3}{c|}{$Cor(T,W)=0.7$} & \multicolumn{3}{c}{$T=W$} \\
                				\multicolumn{4}{c|}{\multirow{2}{*}{}}		& \multicolumn{3}{c|}{$n_j$} &  \multicolumn{3}{c|}{ $n_j$} & \multicolumn{3}{c}{$n_j$} \\
				&&&		& \multicolumn{1}{c}{$30$} & \multicolumn{1}{c}{$40$} & \multicolumn{1}{c|}{$50$} & \multicolumn{1}{c}{$30$} & \multicolumn{1}{c}{$40$} & \multicolumn{1}{c|}{$50$} & \multicolumn{1}{c}{$30$} & \multicolumn{1}{c}{$40$} & \multicolumn{1}{c}{$50$} \\
                				\hline \hline
				\parbox[r]{2mm}{\multirow{6}{*}{\rotatebox[origin=r]{90}{{Bias}}}} &\parbox[r]{2mm}{\multirow{2}{*}{\rotatebox[origin=r]{90}{\small{$r=0$}}}}&\parbox[r]{2mm}{\multirow{2}{*}{\rotatebox[origin=r]{90}{{$M$}}}}&$30$ & -.008 & -.008 & -.005 & -.004 & -.004 & -.002 & ${\tiny{<}}\lvert\!.001\!\rvert$ & ${\tiny{<}}\lvert\!.001\!\rvert$ & .001\\
				&&&$50$ & -.006 & -.005 & -.004 & -.003 & -.002 & -.002 & ${\tiny{<}}\lvert\!.001\!\rvert$ & ${\tiny{<}}\lvert\!.001\!\rvert$ & ${\tiny{<}}\lvert\!.001\!\rvert$\\
				\cline{2-13}
				&\parbox[r]{2mm}{\multirow{2}{*}{\rotatebox[origin=r]{90}{\small{$r=.2$}}}}&\parbox[r]{2mm}{\multirow{2}{*}{\rotatebox[origin=r]{90}{{$M$}}}}&$30$ & -.005 & -.009 & -.007 & -.002 & -.004 & -.004 & .001 & ${\tiny{<}}\lvert\!.001\!\rvert$ & -.001\\
				&&&$50$ & -.007 & -.006 & -.005 & -.003 & -.003 & -.002 & .002 & .001 & .001\\
				\cline{2-13}
				&\parbox[r]{2mm}{\multirow{2}{*}{\rotatebox[origin=r]{90}{\small{$r=.4$}}}}&\parbox[r]{2mm}{\multirow{2}{*}{\rotatebox[origin=r]{90}{{$M$}}}}&$30$ & -.008 & -.009 & -.008 & -.003 & -.005 & -.005 & .002 & -.001 & -.002\\
				&&&$50$ & -.005 & -.005 & -.005 & -.002 & -.002 & -.002 & .002 & .001 & .001\\
				\hline \hline
				\parbox[r]{2mm}{\multirow{6}{*}{\rotatebox[origin=r]{90}{{Standard  deviation}}}}&\parbox[r]{2mm}{\multirow{2}{*}{\rotatebox[origin=r]{90}{\small{$r=0$}}}}&\parbox[r]{2mm}{\multirow{2}{*}{\rotatebox[origin=r]{90}{{$M$}}}}&$30$ & .035 & .031 & .028 & .034 & .030 & .027 & .035 & .031 & .028\\
				&&&$50$ & .028 & .024 & .021 & .027 & .023 & .021 & .028 & .024 & .021\\
				\cline{2-13}
				&\parbox[r]{2mm}{\multirow{2}{*}{\rotatebox[origin=r]{90}{\small{$r=.2$}}}}&\parbox[r]{2mm}{\multirow{2}{*}{\rotatebox[origin=r]{90}{{$M$}}}}&$30$ & .037 & .032 & .029 & .036 & .031 & .028 & .037 & .032 & .029\\
				&&&$50$ & .028 & .024 & .021 & .027 & .023 & .021 & .028 & .024 & .021\\
				\cline{2-13}
				&\parbox[r]{2mm}{\multirow{2}{*}{\rotatebox[origin=r]{90}{\small{$r=.4$}}}}&\parbox[r]{2mm}{\multirow{2}{*}{\rotatebox[origin=r]{90}{{$M$}}}}&$30$ & .043 & .037 & .033 & .042 & .036 & .032 & .043 & .037 & .033\\
				&&&$50$ & .033 & .029 & .026 & .033 & .028 & .025 & .033 & .029 & .026\\
				\hline \hline
				\parbox[r]{2mm}{\multirow{6}{*}{\rotatebox[origin=r]{90}{{Coverage}}}}&\parbox[r]{2mm}{\multirow{2}{*}{\rotatebox[origin=r]{90}{\small{$r=0$}}}}&\parbox[r]{2mm}{\multirow{2}{*}{\rotatebox[origin=r]{90}{{$M$}}}}&$30$ & 93.5 & 93{.0} & 92.9 & 93.7 & 93.6 & 92.2 & 93.8 & 93.8 & 92.2\\
				&&&$50$ & 93.2 & 94.7 & 94.6 & 94.2 & 95.3 & 94.9 & 94.2 & 94.5 & 94.1\\
				\cline{2-13}
				&\parbox[r]{2mm}{\multirow{2}{*}{\rotatebox[origin=r]{90}{\small{$r=.2$}}}}&\parbox[r]{2mm}{\multirow{2}{*}{\rotatebox[origin=r]{90}{{$M$}}}}&$30$ & 83.5 & 77.3 & 77.8 & 83.2 & 78.6 & 76.9 & 83.7 & 79.9 & 77.7\\
				&&&$50$ & 81.4 & 81.2 & 77.9 & 82.4 & 80.6 & 76.8 & 83.2 & 80.8 & 77.3\\
				\cline{2-13}
				&\parbox[r]{2mm}{\multirow{2}{*}{\rotatebox[origin=r]{90}{\small{$r=.4$}}}}&\parbox[r]{2mm}{\multirow{2}{*}{\rotatebox[origin=r]{90}{{$M$}}}}&$30$ & 66.7 & 61.2 & 60.1 & 66.6 & 60.6 & 58.5 & 67.6 & 61{.0} & 60.3\\
				&&&$50$ & 65.7 & 65.2 & 58.5 & 65.2 & 63.1 & 56.3 & 66.1 & 62.6 & 55.2\\
				\hline
		\end{tabular}}
	\end{table}

	\begin{table}[!htb]
		\centering
		{\small	\caption{
        Summary of the estimates (bias, standard deviation, coverage) of $\tau$ over 1,000 Monte Carlo replicates and 1,000 bootstrap samples for the regular Bayesian bootstrap with propensity score regression marginalized model under a multilevel data-generating process. Results for variations in $r$, $n_j$, and $M$. Covariate summary via propensity score adjustment.
        }
			\label{tab:rescausal2ml}
            			\renewcommand{\arraystretch}{1.3}
                        			\begin{tabular}{llll|rrr|rrr|rrr}
				\hline
				\multicolumn{4}{c|}{\multirow{3}{*}{Reg. BB-ML}}		& \multicolumn{3}{c|}{$Cor(T,W)=0$} &  \multicolumn{3}{c|}{$Cor(T,W)=0.7$} & \multicolumn{3}{c}{$T=W$} \\
                				\multicolumn{4}{c|}{\multirow{2}{*}{}}		& \multicolumn{3}{c|}{$n_j$} &  \multicolumn{3}{c|}{ $n_j$} & \multicolumn{3}{c}{$n_j$} \\
				&&&		& \multicolumn{1}{c}{$30$} & \multicolumn{1}{c}{$40$} & \multicolumn{1}{c|}{$50$} & \multicolumn{1}{c}{$30$} & \multicolumn{1}{c}{$40$} & \multicolumn{1}{c|}{$50$} & \multicolumn{1}{c}{$30$} & \multicolumn{1}{c}{$40$} & \multicolumn{1}{c}{$50$} \\
                				\hline \hline
				\parbox[r]{2mm}{\multirow{6}{*}{\rotatebox[origin=r]{90}{{Bias}}}} &\parbox[r]{2mm}{\multirow{2}{*}{\rotatebox[origin=r]{90}{\small{$r=0$}}}}&\parbox[r]{2mm}{\multirow{2}{*}{\rotatebox[origin=r]{90}{{$M$}}}}&$30$ & -.001 & .005 & -.004 & .471 & .438 & .471 & .952 & .880 & .955\\
				&&&$50$ & .001 & .001 & ${\tiny{<}}\lvert\!.001\!\rvert$ & .492 & .458 & .453 & .994 & .924 & .915\\
				\cline{2-13}
				&\parbox[r]{2mm}{\multirow{2}{*}{\rotatebox[origin=r]{90}{\small{$r=.2$}}}}&\parbox[r]{2mm}{\multirow{2}{*}{\rotatebox[origin=r]{90}{{$M$}}}}&$30$ & .002 & ${\tiny{<}}\lvert\!.001\!\rvert$ & -.002 & .450 & .445 & .476 & .906 & .899 & .964\\
				&&&$50$ & .002 & -.003 & -.001 & .423 & .472 & .470 & .853 & .956 & .951\\
				\cline{2-13}
				&\parbox[r]{2mm}{\multirow{2}{*}{\rotatebox[origin=r]{90}{\small{$r=.4$}}}}&\parbox[r]{2mm}{\multirow{2}{*}{\rotatebox[origin=r]{90}{{$M$}}}}&$30$ & .008 & -.002 & -.002 & .477 & .443 & .476 & .956 & .897 & .964\\
				&&&$50$ & .003 & -.005 & -.002 & .477 & .469 & .469 & .961 & .953 & .950\\
				\hline \hline
				\parbox[r]{2mm}{\multirow{6}{*}{\rotatebox[origin=r]{90}{{Standard  deviation}}}}&\parbox[r]{2mm}{\multirow{2}{*}{\rotatebox[origin=r]{90}{\small{$r=0$}}}}&\parbox[r]{2mm}{\multirow{2}{*}{\rotatebox[origin=r]{90}{{$M$}}}}&$30$ & .034 & .027 & .026 & .020 & .017 & .015 & .017 & .015 & .013\\
				&&&$50$ & .027 & .022 & .019 & .016 & .012 & .011 & .015 & .010 & .009\\
				\cline{2-13}
				&\parbox[r]{2mm}{\multirow{2}{*}{\rotatebox[origin=r]{90}{\small{$r=.2$}}}}&\parbox[r]{2mm}{\multirow{2}{*}{\rotatebox[origin=r]{90}{{$M$}}}}&$30$ & .034 & .030 & .028 & .022 & .019 & .018 & .019 & .018 & .015\\
				&&&$50$ & .025 & .024 & .022 & .017 & .016 & .014 & .016 & .014 & .013\\
				\cline{2-13}
				&\parbox[r]{2mm}{\multirow{2}{*}{\rotatebox[origin=r]{90}{\small{$r=.4$}}}}&\parbox[r]{2mm}{\multirow{2}{*}{\rotatebox[origin=r]{90}{{$M$}}}}&$30$ & .040 & .034 & .032 & .030 & .026 & .024 & .027 & .025 & .022\\
				&&&$50$ & .033 & .028 & .025 & .024 & .021 & .019 & .023 & .020 & .018\\
				\hline \hline
				\parbox[r]{2mm}{\multirow{6}{*}{\rotatebox[origin=r]{90}{{Coverage}}}}&\parbox[r]{2mm}{\multirow{2}{*}{\rotatebox[origin=r]{90}{\small{$r=0$}}}}&\parbox[r]{2mm}{\multirow{2}{*}{\rotatebox[origin=r]{90}{{$M$}}}}&$30$ & 26.9 & 22.7 & 20.7 & .0 & .0 & .0 & .0 & .0 & .0\\
				&&&$50$ & 23.2 & 22.7 & 20.2 & .0 & .0 & .0 & .0 & .0 & .0\\
				\cline{2-13}
				&\parbox[r]{2mm}{\multirow{2}{*}{\rotatebox[origin=r]{90}{\small{$r=.2$}}}}&\parbox[r]{2mm}{\multirow{2}{*}{\rotatebox[origin=r]{90}{{$M$}}}}&$30$ & 27.1 & 19.9 & 20.7 & .1 & .2 & .0 & .0 & .0 & .0\\
				&&&$50$ & 25.9 & 23.2 & 20.9 & .0 & .0 & .0 & .0 & .0 & .0\\
				\cline{2-13}
				&\parbox[r]{2mm}{\multirow{2}{*}{\rotatebox[origin=r]{90}{\small{$r=.4$}}}}&\parbox[r]{2mm}{\multirow{2}{*}{\rotatebox[origin=r]{90}{{$M$}}}}&$30$ & 25.2 & 22.1 & 21.5 & .7 & .9 & .9 & .0 & .0 & .0\\
				&&&$50$ & 27.2 & 24.1 & 20.2 & .2 & .2 & .0 & .0 & .0 & .0\\
				\hline
		\end{tabular}}
	\end{table}

	\begin{table}[!htb]
		\centering
		{\small	\caption{
        Summary of the estimates (bias, standard deviation, coverage) of $\tau$ over 1,000 Monte Carlo replicates and 1,000 bootstrap samples for the EDW-BB with propensity score regression marginalized model under a multilevel data-generating process. Results for variations in $r$, $n_j$, and $M$. Covariate summary via propensity score adjustment. }
			\label{tab:rescausal3ml}
			\renewcommand{\arraystretch}{1.3}
                                    			\begin{tabular}{llll|rrr|rrr|rrr}
				\hline
				\multicolumn{4}{c|}{\multirow{3}{*}{EDW-BB-ML}}		& \multicolumn{3}{c|}{$Cor(T,W)=0$} &  \multicolumn{3}{c|}{$Cor(T,W)=0.7$} & \multicolumn{3}{c}{$T=W$} \\
                				\multicolumn{4}{c|}{\multirow{2}{*}{}}		& \multicolumn{3}{c|}{$n_j$} &  \multicolumn{3}{c|}{ $n_j$} & \multicolumn{3}{c}{$n_j$} \\
				&&&		& \multicolumn{1}{c}{$30$} & \multicolumn{1}{c}{$40$} & \multicolumn{1}{c|}{$50$} & \multicolumn{1}{c}{$30$} & \multicolumn{1}{c}{$40$} & \multicolumn{1}{c|}{$50$} & \multicolumn{1}{c}{$30$} & \multicolumn{1}{c}{$40$} & \multicolumn{1}{c}{$50$} \\
                				\hline \hline
				\parbox[r]{2mm}{\multirow{6}{*}{\rotatebox[origin=r]{90}{{Bias}}}} &\parbox[r]{2mm}{\multirow{2}{*}{\rotatebox[origin=r]{90}{\small{$r=0$}}}}&\parbox[r]{2mm}{\multirow{2}{*}{\rotatebox[origin=r]{90}{{$M$}}}}&$30$ & ${\tiny{<}}\lvert\!.001\!\rvert$ & .006 & -.003 & .467 & .434 & .469 & .944 & .872 & .951\\
				&&&$50$ & .002 & .002 & ${\tiny{<}}\lvert\!.001\!\rvert$ & .492 & .458 & .451 & .991 & .924 & .912\\
				\cline{2-13}
				&\parbox[r]{2mm}{\multirow{2}{*}{\rotatebox[origin=r]{90}{\small{$r=.2$}}}}&\parbox[r]{2mm}{\multirow{2}{*}{\rotatebox[origin=r]{90}{{$M$}}}}&$30$ & .003 & .001 & -.002 & .448 & .441 & .475 & .902 & .890 & .961\\
				&&&$50$ & .003 & -.002 & -.001 & .422 & .470 & .468 & .851 & .951 & .947\\
				\cline{2-13}
				&\parbox[r]{2mm}{\multirow{2}{*}{\rotatebox[origin=r]{90}{\small{$r=.4$}}}}&\parbox[r]{2mm}{\multirow{2}{*}{\rotatebox[origin=r]{90}{{$M$}}}}&$30$ & .008 & -.001 & -.002 & .475 & .439 & .475 & .952 & .889 & .961\\
				&&&$50$ & .004 & -.004 & -.001 & .476 & .467 & .468 & .957 & .948 & .947\\
				\hline \hline
				\parbox[r]{2mm}{\multirow{6}{*}{\rotatebox[origin=r]{90}{{Standard  deviation}}}}&\parbox[r]{2mm}{\multirow{2}{*}{\rotatebox[origin=r]{90}{\small{$r=0$}}}}&\parbox[r]{2mm}{\multirow{2}{*}{\rotatebox[origin=r]{90}{{$M$}}}}&$30$ & .167 & .157 & .168 & .091 & .088 & .091 & .066 & .070 & .064\\
				&&&$50$ & .138 & .127 & .125 & .077 & .068 & .068 & .066 & .046 & .047\\
				\cline{2-13}
				&\parbox[r]{2mm}{\multirow{2}{*}{\rotatebox[origin=r]{90}{\small{$r=.2$}}}}&\parbox[r]{2mm}{\multirow{2}{*}{\rotatebox[origin=r]{90}{{$M$}}}}&$30$ & .169 & .170 & .178 & .103 & .106 & .108 & .079 & .092 & .083\\
				&&&$50$ & .127 & .140 & .140 & .080 & .087 & .087 & .069 & .072 & .075\\
				\cline{2-13}
				&\parbox[r]{2mm}{\multirow{2}{*}{\rotatebox[origin=r]{90}{\small{$r=.4$}}}}&\parbox[r]{2mm}{\multirow{2}{*}{\rotatebox[origin=r]{90}{{$M$}}}}&$30$ & .199 & .195 & .204 & .142 & .143 & .146 & .122 & .133 & .129\\
				&&&$50$ & .163 & .163 & .163 & .119 & .120 & .120 & .106 & .109 & .111\\
				\hline \hline
				\parbox[r]{2mm}{\multirow{6}{*}{\rotatebox[origin=r]{90}{{Coverage}}}}&\parbox[r]{2mm}{\multirow{2}{*}{\rotatebox[origin=r]{90}{\small{$r=0$}}}}&\parbox[r]{2mm}{\multirow{2}{*}{\rotatebox[origin=r]{90}{{$M$}}}}&$30$ & 89.7 & 89.9 & 89.8 & .7 & 1.1 & .4 & .0 & .0 & .0\\
				&&&$50$ & 92.0 & 92.1 & 92.0 & .0 & .0 & .0 & .0 & .0 & .0\\
				\cline{2-13}
				&\parbox[r]{2mm}{\multirow{2}{*}{\rotatebox[origin=r]{90}{\small{$r=.2$}}}}&\parbox[r]{2mm}{\multirow{2}{*}{\rotatebox[origin=r]{90}{{$M$}}}}&$30$ & 89.1 & 89.4 & 89.1 & 2.4 & 3.8 & 2.6 & .0 & .0 & .0\\
				&&&$50$ & 91.5 & 91.3 & 90.5 & .3 & .1 & .3 & .0 & .0 & .0\\
				\cline{2-13}
				&\parbox[r]{2mm}{\multirow{2}{*}{\rotatebox[origin=r]{90}{\small{$r=.4$}}}}&\parbox[r]{2mm}{\multirow{2}{*}{\rotatebox[origin=r]{90}{{$M$}}}}&$30$ & 88.4 & 88.5 & 87.5 & 13.4 & 17.8 & 15.6 & .3 & .2 & .0\\
				&&&$50$ & 91.7 & 91.0 & 90.2 & 5.2 & 5.9 & 6.7 & .0 & .0 & .0\\
				\hline
		\end{tabular}}
	\end{table}
	\begin{table}[!htb]
		\centering
		{\small	\caption{
        Summary of the estimates (bias, standard deviation, coverage) of $\tau$ over 1,000 Monte Carlo replicates and 1,000 bootstrap samples for the unweighted  model under a multilevel data-generating process. Results for variations in $r$, $n_j$, and $M$. Additional results of coverage rates for $M \in \{30,50\}$ and $n_j \in \{80,100,120\}$. Covariate summary via propensity score adjustment.  }
			\label{tab:rescausal2ext}
			\renewcommand{\arraystretch}{1.3}
                                               			\begin{tabular}{llll|rrr|rrr|rrr}
				\hline
				\multicolumn{4}{c|}{\multirow{3}{*}{Unw+PSadj}}		& \multicolumn{3}{c|}{$Cor(T,W)=0$} &  \multicolumn{3}{c|}{$Cor(T,W)=0.7$} & \multicolumn{3}{c}{$T=W$} \\
                				\multicolumn{4}{c|}{\multirow{2}{*}{}}		& \multicolumn{3}{c|}{$n_j$} &  \multicolumn{3}{c|}{ $n_j$} & \multicolumn{3}{c}{$n_j$} \\
				&&&		& \multicolumn{1}{c}{$30$} & \multicolumn{1}{c}{$40$} & \multicolumn{1}{c|}{$50$} & \multicolumn{1}{c}{$30$} & \multicolumn{1}{c}{$40$} & \multicolumn{1}{c|}{$50$} & \multicolumn{1}{c}{$30$} & \multicolumn{1}{c}{$40$} & \multicolumn{1}{c}{$50$} \\
                				\hline  \hline
				\parbox[r]{2mm}{\multirow{6}{*}{\rotatebox[origin=r]{90}{{Coverage: 60\%}}}} &\parbox[r]{2mm}{\multirow{2}{*}{\rotatebox[origin=r]{90}{{$r=0$}}}}&\parbox[r]{2mm}{\multirow{2}{*}{\rotatebox[origin=r]{90}{{$M$}}}}&$30$ & 99.9 & 100{.0} & 100{.0} & 97.2 & 98.4 & 96.9 & 80{.0} & 83.4 & 82.1\\
				&&&$50$ & 100{.0} & 100{.0} & 100{.0} & 99.4 & 98.3 & 98.6 & 85.2 & 84{.0} & 81.8\\
				\cline{2-13}
				&\parbox[r]{2mm}{\multirow{2}{*}{\rotatebox[origin=r]{90}{{$r=.2$}}}}&\parbox[r]{2mm}{\multirow{2}{*}{\rotatebox[origin=r]{90}{{$M$}}}}&$30$ & 96.2 & 96.1 & 93.3 & 79.4 & 79{.0} & 74.2 & 62.1 & 62.4 & 54.9\\
				&&&$50$ & 99.2 & 97.4 & 95.6 & 85{.0} & 81.2 & 79.2 & 67.1 & 61.3 & 59.3\\
				\cline{2-13}
				&\parbox[r]{2mm}{\multirow{2}{*}{\rotatebox[origin=r]{90}{{$r=.4$}}}}&\parbox[r]{2mm}{\multirow{2}{*}{\rotatebox[origin=r]{90}{{$M$}}}}&$30$ & 80.5 & 80.8 & 74.5 & 61.7 & 62.3 & 55.1 & 52{.0} & 51{.0} & 46.3\\
				&&&$50$ & 86.8 & 81.4 & 79{.0} & 69.1 & 62.2 & 60{.0} & 58.8 & 53{.0} & 49{.0}\\
				\hline  \hline
				\parbox[r]{2mm}{\multirow{6}{*}{\rotatebox[origin=r]{90}{{Coverage: 80\%}}}}&\parbox[r]{2mm}{\multirow{2}{*}{\rotatebox[origin=r]{90}{{$r=0$}}}}&\parbox[r]{2mm}{\multirow{2}{*}{\rotatebox[origin=r]{90}{{$M$}}}}&$30$  & 100{.0} & 100{.0} & 100{.0} & 99.9 & 100{.0} & 99.9 & 93.4 & 96.3 & 94.1\\
				&&&$50$ & 100{.0} & 100{.0} & 100{.0} & 100{.0} & 100{.0} & 100{.0} & 96.4 & 96.5 & 95.8\\
				\cline{2-13}
				&\parbox[r]{2mm}{\multirow{2}{*}{\rotatebox[origin=r]{90}{{$r=.2$}}}}&\parbox[r]{2mm}{\multirow{2}{*}{\rotatebox[origin=r]{90}{{$M$}}}}&$30$ & 99.8 & 99.6 & 99{.0} & 92.9 & 92.1 & 89.7 & 78.9 & 78.6 & 73.9\\
				&&&$50$ & 99.9 & 100{.0} & 99.5 & 97.3 & 95.1 & 93.3 & 84.2 & 80.7 & 78{.0}\\
				\cline{2-13}
				&\parbox[r]{2mm}{\multirow{2}{*}{\rotatebox[origin=r]{90}{{$r=.4$}}}}&\parbox[r]{2mm}{\multirow{2}{*}{\rotatebox[origin=r]{90}{{$M$}}}}&$30$ & 93{.0} & 92.9 & 88.7 & 78.6 & 79.9 & 72.9 & 69.3 & 69.9 & 62.3\\
				&&&$50$ & 98.3 & 95.4 & 93.1 & 85.4 & 80.7 & 79.7 & 76.4 & 70.3 & 68.7\\
				\hline  \hline
				\parbox[r]{2mm}{\multirow{6}{*}{\rotatebox[origin=r]{90}{{Coverage: 95\%}}}}&\parbox[r]{2mm}{\multirow{2}{*}{\rotatebox[origin=r]{90}{{$r=0$}}}}&\parbox[r]{2mm}{\multirow{2}{*}{\rotatebox[origin=r]{90}{{$M$}}}}&$30$ & 100{.0} & 100{.0} & 100{.0} & 100{.0} & 100{.0} & 100{.0} & 99.3 & 99.8 & 99.8\\
				&&&$50$ & 100{.0} & 100{.0} & 100{.0} & 100{.0} & 100{.0} & 100{.0} & 99.9 & 99.7 & 99.7\\
				\cline{2-13}
				&\parbox[r]{2mm}{\multirow{2}{*}{\rotatebox[origin=r]{90}{{$r=.2$}}}}&\parbox[r]{2mm}{\multirow{2}{*}{\rotatebox[origin=r]{90}{{$M$}}}}&$30$ & 100{.0} & 100{.0} & 100{.0} & 99.4 & 99.2 & 98.5 & 92.7 & 93.6 & 90.3\\
				&&&$50$ & 100{.0} & 100{.0} & 100{.0} & 99.8 & 99.6 & 98.3 & 96.8 & 95.4 & 92.6\\
				\cline{2-13}
				&\parbox[r]{2mm}{\multirow{2}{*}{\rotatebox[origin=r]{90}{{$r=.4$}}}}&\parbox[r]{2mm}{\multirow{2}{*}{\rotatebox[origin=r]{90}{{$M$}}}}&$30$ & 99.3 & 99.6 & 98{.0} & 94.1 & 92.8 & 90{.0} & 87.1 & 85.8 & 81.9\\
				&&&$50$ & 99.7 & 99.8 & 98.8 & 98.2 & 96.2 & 92.9 & 92.5 & 89.9 & 86.9\\
				\hline
		\end{tabular}}
	\end{table}

\clearpage
	
	\section{Supplementary Material: Plasmode Simulation set-up - tuberculosis data}\label{app:plamode}
	
\begin{figure}[!htb]
		\centering
		\includegraphics[scale=.3]{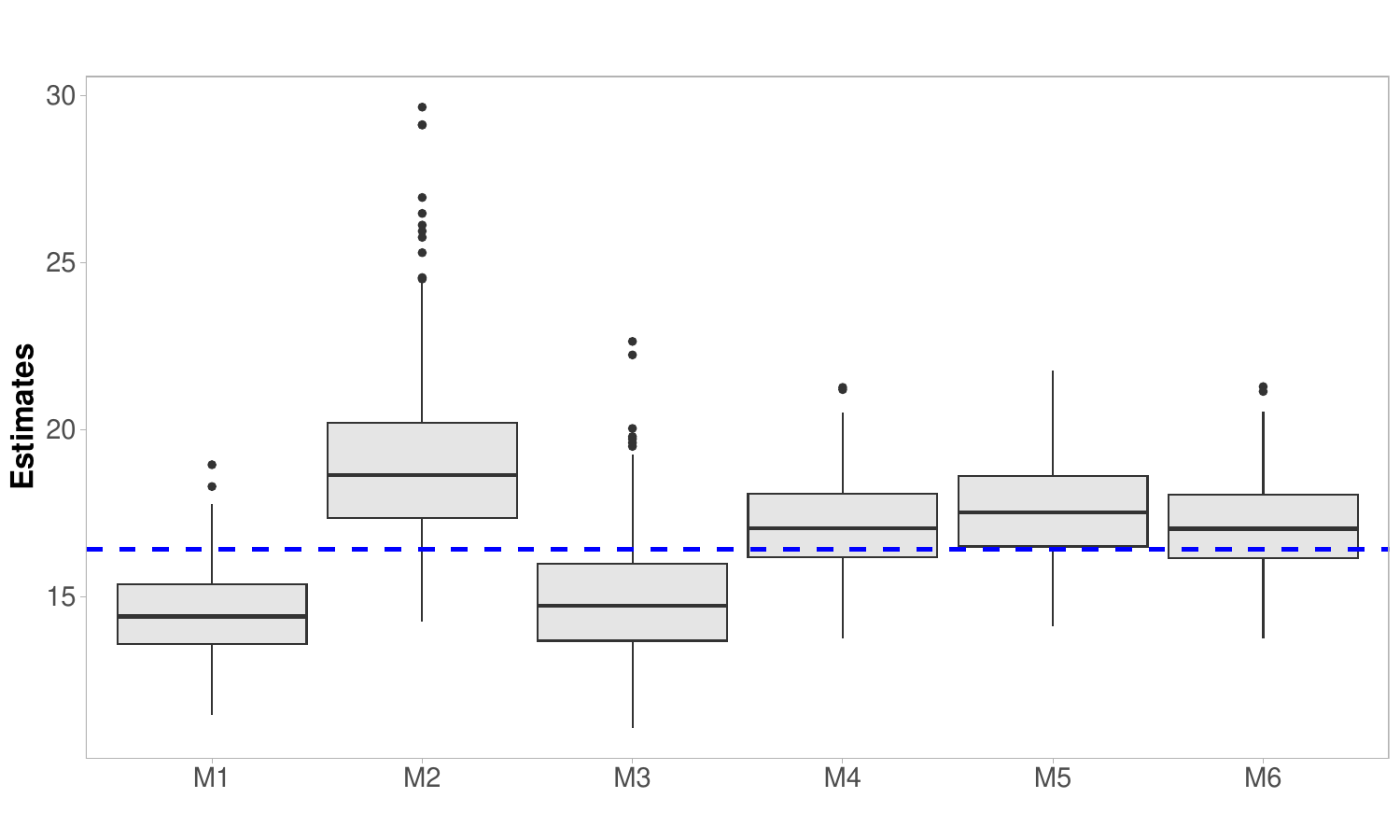}
		\caption{Boxplot of the posterior means of the \textit{population} odds ratio of the plasmode Simulation set-up of the TB data. The horizontal dashed line denotes the true value used in the simulation. }
		\label{fig:tb-res-plas}
	\end{figure}

	This section provides a plasmode Simulation set-up of the tuberculosis dataset to clarify the results presented in Section \ref{sec:psr} of the main file. The data-generating process uses the true $Z$ and $X$ values as observed data and simulates $Y$ according to the following model:
	\[
	\begin{split}
		Y_{ji}|\hat{\mu}_{ji} & \sim Bernoulli (\hat{\mu}_{ji}),
	\end{split}
	\]
	where $\hat{\mu}_{ji}$ denotes the fitted values obtained from the model
	\[
	\begin{split}
		{\rm logit}\left({\mu}_{ji}\right) & = (\beta_0 + {\lambda}_{0j}) + (\beta_1 + {\lambda}_{1j})Z_{ji} + \mathbf{X}^{\top}_{ji}{\boldsymbol{\delta}}.
	\end{split}
	\]
	This specification corresponds to a logistic regression model with random intercept and slope, fitted using the \texttt{stan\_glmer} function in the R package \texttt{rstanarm} (Goodrich et al., 2025). A total of 400 data replicates of $Y$ were generated, and the models from Table \ref{tab:tb-md} in the main text were fitted accordingly. The true odds ratio (OR) from the data-generating process is 16.64, a value chosen to mimic the estimates found in the real data. The posterior mean standard deviations for the intercept and $Z$-slope random effects were estimated at 0.52 and 0.45, respectively. 
	Table \ref{tab:tb-md-plasmode} presents metrics of comparison for the fitted models, while Figure \ref{fig:tb-res-plas} displays boxplots of the posterior estimates across the 400 data replicates for each model. Additionally, a boxplot of the posterior distribution of the \textit{population} OR used in the plasmode data-generating process is included for reference. The abbreviation APoM refers to the average posterior mean over 400 data replicates, and rMSE denotes the root mean square error.
		\begin{table}[!htb]
		\centering
		\caption{Summary of the \textit{population} OR for each model described in Table \ref{tab:tb-md}, of the main text, for the plasmode Simulation set-up. }
		\label{tab:tb-md-plasmode}
		\begin{tabular}{l|rrrr}
			\hline
			\multirow{2}{*}{Model} & \multicolumn{1}{c}{\multirow{2}{*}{APoM}} & \multicolumn{1}{c}{\multirow{2}{*}{Bias}} &  \multicolumn{1}{c}{\multirow{2}{*}{rMSE}} & \multicolumn{1}{c}{Coverage} \\
			& & & & \multicolumn{1}{c}{(95\%)} \\
			\hline
			M1 &14.52 & -2.12 & 2.85 & 65.8\\ 
			M2 & 19.04 & 2.40 & 4.51 & 90.5\\ 
			M3 & 14.95 & -1.69& 3.37 & 90.2\\ 
			M4 &17.17 & 0.53& 2.16 &  96.5\\ 
			M5 & 17.16 & 0.52 &2.15 & 96{.0}\\
			M6 & 17.64 & 1.00 & 3.42 & 100{.0}\\ 
			\hline
		\end{tabular}
	\end{table}
	Results from 400 data replicates indicate that all models considered produce biased estimates. Models M4 and M5 outperform the others across all metrics, including coverage rates that are closer to the nominal level. These findings contrast with those of the previous Simulation set-up, which suggest that M1 may maintain nominal coverage rates for specific combinations of $M$, $n_j$, and the standard deviation of the slope random effect in the data-generating process. However, coverage decreases as $M$ and/or $n_j$ increase. In contrast, models incorporating random effects tend to be biased and exhibit poor coverage. Model M6, in particular, shows evidence of bias and overestimation of the posterior variance, with a coverage rate of $100\%$. The boxplots of the point estimates further demonstrate that including random effects can substantially alter posterior summaries.

	\bibliography{BibReferences_BA_revised}
	\bibliographystyle{ba}

\end{document}